\documentclass[a4paper,11pt]{amsproc}

\usepackage{amsmath}
\usepackage{bbm}
\usepackage{cancel}
\usepackage[normalem]{ulem}
\usepackage{citesort}
\usepackage[sort&compress,numbers]{natbib}

\usepackage{amsfonts}
\usepackage{amssymb}
\usepackage{amsmath}
\usepackage{latexsym}
\usepackage{xcolor}
\usepackage{url}

\newcommand{\dan}{\mbox{$ \parallel \mkern -17mu - \ $}}

 \newcommand{\lalpha}{{\tilde \alpha}}

\newcommand{\mydivergence}{{\mathrm{div}}}

\newcommand{\curl}{{\mathrm{curl}}}
\newcommand{\trace}{{\mathrm{tr}}}

\newcommand{\mtot}{{m_{\mathrm{total}}}}

\newcommand{\rfcp}{{retarded foliation near $i^0$}}
\newcommand{\rfscp}{{retarded foliations near $i^0$}}

\newcommand{\normwithnodelta }{|}

\newcommand{\bMinkowskimetric}{\eta}
\newcommand{\hnews}{N}

\newcommand{\mTB}{m_{\mbox{\rm\scriptsize TB}} }

\newcommand{\nlap}{\mbox{$ \nabla \mkern -13mu / \ $}}

\newcounter{mnotecount}[section]

\renewcommand{\themnotecount}{\thesection.\arabic{mnotecount}}
\newcommand{\mnote}[1]
{\protect{\stepcounter{mnotecount}}$^{\mbox{\footnotesize
$
\bullet$\themnotecount}}$ \marginpar{
\raggedright\tiny\em
$\!\!\!\!\!\!\,\bullet$\themnotecount: #1} }

\newtheorem{theorem}{\sc  Theorem\rm}[section]

\newtheorem{Theorem}[theorem]{\sc  Theorem\rm}
\newtheorem{The}[theorem]{\sc  Theorem\rm}

\newtheorem{Def}[theorem]{\sc  Definition\rm}

\newtheorem{proposition}[theorem]{\sc Proposition\rm}
\newtheorem{prop}[theorem]{\sc Proposition\rm}
\newtheorem{Proposition}[theorem]{\sc Proposition\rm}

\newtheorem{Remark}[theorem]{\sc Remark\rm}

\newcommand{\ol}[1]{\overline{#1}{}}

\newcommand{\jlcax}[1]{}
\newcommand{\eean}{\nonumber\end{eqnarray}}

\newcommand{\og}{{\overline{g}}}

\newcommand{\fourg}{{\mathfrak g }}

\newcommand{\kk}[1]{}

\newcommand{\beq}{\begin{equation}}

\newcommand{\rd}{\,{ d}} 

\newcommand{\FS}       
                  {F}

\newcommand{\HS} 
       {H_{\mbox{\scriptsize volume}}}

{\ptc{this should be removed in the oberwolfach version}}%

\newcommand{\ourU}{\mathbb U}%
\newcommand{\eeal}[1]{\label{#1}\end{eqnarray}}
\newcommand{\bed}{\begin{deqarr}}
\newcommand{\eed}{\end{deqarr}}
\newcommand{\bedl}[1]{\begin{deqarr}\label{#1}}
\newcommand{\eedl}[2]{\arrlabel{#1}\label{#2}\end{deqarr}}

\newcommand{\mcN}{{\mycal N}}

\newcommand{\bel}[1]{\begin{equation}\label{#1}}
\newcommand{\bea}{\begin{eqnarray}}
\newcommand{\bean}{\begin{eqnarray}\nonumber}
\newcommand{\beal}[1]{\begin{eqnarray}\label{#1}}
\newcommand{\eea}{\end{eqnarray}}

\newcommand{\nn}{\nonumber}

\newcommand{\be}{\begin{equation}}
\newcommand{\eeq}{\end{equation}}
\newcommand{\ee}{\end{equation}}
\newcommand{\beqa}{\begin{eqnarray}}
\newcommand{\eeqa}{\end{eqnarray}}
\newcommand{\beqan}{\begin{eqnarray*}}
\newcommand{\eeqan}{\end{eqnarray*}}
\newcommand{\ba}{\begin{array}}
\newcommand{\ea}{\end{array}}

\newcommand{\hyp}{\mycal S}

\newcommand{\mcM}{{\mycal M}}

\newcommand{\scri}{{\mycal I}}%
\newcommand{\scrip}{\scri^{+}}%
\newcommand{\Scri}{\scri}

\newcommand{\warn}[1]
{\protect{\stepcounter{mnotecount}}$^{\mbox{\footnotesize
$
\bullet$\themnotecount}}$ \marginpar{
\raggedright\tiny\em
$\!\!\!\!\!\!\,\bullet$\themnotecount: {\bf Warning:} #1} }

\newcommand{\R}{\mathbb R}

\newcommand{\N}{\mathbb N}

\newcommand{\eq}[1]{(\ref{#1})}

\newcommand{\ptc}[1]{\mnote{{\bf ptc:}#1}}

\newcommand{\mcS}{{\mycal S}}
\newcommand{\mcL}{{\mycal L}}

\newcommand{\beqar}{\begin{deqarr}}
\newcommand{\eeqar}{\end{deqarr}}

\newcommand{\beaa}{\begin{eqnarray*}}
\newcommand{\eeaa}{\end{eqnarray*}}

\newcommand{\tr}{\mbox{tr}}

\newcommand{\calM}{{\mcM}}

\renewcommand{\fourg}{{}^4g}

\DeclareFontFamily{OT1}{rsfs}{}
\DeclareFontShape{OT1}{rsfs}{m}{n}{ <-7> rsfs5 <7-10> rsfs7 <10-> rsfs10}{}
\DeclareMathAlphabet{\mycal}{OT1}{rsfs}{m}{n}

{\catcode `\@=11 \global\let\AddToReset=\@addtoreset}
\AddToReset{equation}{section}

{\catcode `\@=11 \global\let\AddToReset=\@addtoreset}
\AddToReset{figure}{section}

{\catcode `\@=11 \global\let\AddToReset=\@addtoreset}
\AddToReset{table}{section}

\begin{document}
\title[Interior gluings and the Trautman-Bondi mass]{Future-complete null hypersurfaces, interior gluings, and the Trautman-Bondi mass}
 \thanks{Preprint UWThPh-2016-24}

\author{Lydia Bieri}
\address{University of Michigan}
\email{lbieri@umich.edu}

\author{Piotr T.\ Chru\'{s}ciel}
\address{University of Vienna and Erwin Schr\"odinger Institute, Boltzmanngasse 5, A 1090 Vienna, Austria}
\email{piotr.chrusciel@univie.ac.at}


\date{}

\maketitle

\begin{abstract}
We  {present the argument} that the past limit of the Trautman-Bondi mass is the ADM mass under weak  hypotheses on the decay of the metric towards spatial infinity, without any smallness conditions on the initial data, assuming well defined energy, momentum, center of mass and angular moment. {Part of the proof consists of} a careful inspection of the proof of stability of Minkowski space-time, which is sketched. This is complemented by an interior gluing result for asymptotically flat initial data with well defined Poincar\'e charges,  which is proved in detail.
\end{abstract}


\section{Introduction}
 \label{s31VII16.1}

There exist two key notions of total energy in general relativity: the Arnowitt-Deser-Misner (ADM) one, calculated on spheres which recede to infinity in spacelike directions, and the Trautman-Bondi (TB) one, calculated on spheres which recede to infinity in null directions. It is thus natural to analyze the relation between those energies.

In fact, this issue has already been addressed under a specific set of conditions in
\cite{AshtekarAshtekarEnergyMomentum}. However, it is still not known how large is the set of space-times which satisfy the hypotheses set forth in that work. Since there has been meanwhile a lot of progress in the understanding of the problems at hand, it is of interest to return to the question, taking into account the new perspectives.

Clearly,  a prerequisite to studying the relation between the ADM and the TB masses is a collection of space-times where one can recede to infinity both in null and spacelike directions. To make things precise, consider a space-time $(\mcM,{}^4g)$ evolving out of initial data $(\hyp,g,K)$ which are asymptotically flat at large distances in the asymptotic regions, so that the ADM four-momentum of  $(\hyp,g,K)$ can be defined. In order to capture the notion of a family of hypersurfaces which behave as the retarded time coordinate $u=t-r$ in Minkowski space-time we proceed as follows: Suppose that $\mcM$ contains a foliation by null hypersurfaces $\mcN(u)$ parameterised by a parameter
$u\in (-\infty,u_0)$, for some $u_0 \le +\infty$. Assume that the hypersurfaces $\mcN(u)$ intersect the asymptotically flat region of $\hyp$ in spheres which, to leading order, are coordinate spheres in manifestly asymptotically Euclidean coordinates on $\hyp$, and which recede to infinity on $\hyp$ as $u$ tends to minus infinity. Assume that the family of spheres so obtained foliates the asymptotic region of $\hyp$, with the future-directed tangents to the generators of $\mcN(u)$ pointing outwards on $\hyp$. Suppose finally that all generators of each $\mcN(u)$ are complete to the future. Such a family of null hypersurfaces will be referred to as \emph{a \rfcp}.
 Assuming moreover that each hypersurface has a well defined TB mass $m(u)$ (cf.\ Section~\ref{s6V16.5} below), the questions then are
\beal{28VIII16.1}
 &&\mbox{$\bullet$ does the limit $\lim_{u
\to -\infty} m(u) $ exist?  and, if so,}
\\
 &&\mbox{$\bullet$ is it  equal to the ADM mass?}
\eeal{28VIII16.2}

One can view null hypersurfaces as being obtained from spacelike ones by an infinite boost.
The  existence of \rfscp\ can then be thought of as \emph{an infinite boost theorem}. We will prove such a theorem below.  {However, to avoid confusion with the already existing terminology associated with the Aichelburg-Sexl metrics, we will not use the \emph{infinite boost} terminology in our context.}

We further note that we do make any  claims on the regularity of the conformally rescaled metric on the ``piece of $\scri$'', the existence of which can be inferred from our construction.

While the \emph{finite boost theorem} has been proved a long time ago~\cite{christodoulou:murchadha}, the existence of future-complete null hypersurfaces
has only been settled so far for weak gravitational fields, or for restricted classes of initial data, or both~\cite{LindbladRodnianski2,KlainermanNicoloBook,BieriJDG,Ch-Kl,BieriZipser1}. For instance, within the class of space-times evolving out of asymptotically flat initial data, such foliations exist

\begin{enumerate}
  \item \label{p28VIII16.1} (obviously, by uniqueness of solutions in domains of dependence)  for  initial data which are stationary at large distances, or
  \item \label{p28VIII16.2}
  for small initial data with \emph{optimal} asymptotic conditions~\cite{BieriJDG} or,
  \item \label{p28VIII16.4}
  without smallness restrictions,
  for CMC initial data which asymptote to   Schwarzschildean data to high order~\cite{KlainermanNicoloBook};
\end{enumerate}
see Section~\ref{s3V16.1} below for a more detailed discussion.  Optimality in point~\ref{p28VIII16.2} is understood with respect to the possibility of defining the ADM mass.

In point~\ref{p28VIII16.1}.\  the TB mass $m(u)$ equals the ADM mass for all large negative $u$, so the answer to both questions  \eq{28VIII16.1}-\eq{28VIII16.2} is immediate. These questions in case~\ref{p28VIII16.2}., in which the asymptotic conditions on the data are the weakest possible compatible with well-defined and finite energy momentum, are open. (In Section~\ref{s2V16.2} below we propose a continuity strategy which could settle this.)
Both questions \eq{28VIII16.1}-\eq{28VIII16.2} have been answered positively in case~\ref{p28VIII16.4}.\ in~\cite{KlainermanNicoloBook}, but the decay conditions imposed on the initial data there are much stronger than desirable.

In this work we  {sketch the proof of}  an affirmative answer to \eq{28VIII16.1}-\eq{28VIII16.2}  for asymptotically flat data with optimal pointwise-weighted  decay conditions on the initial data, without any smallness conditions  assuming, however, a parity condition which guarantees well defined total energy, momentum, center of mass and angular momentum.  {Some elements of the proof are given in detail. The parity condition will be removed in~\cite{ChBoost}.
This, together with  the results in~\cite{Bieriprep},}  results in a statement which is optimal in weighted H\"older spaces, in the sense that all relevant H\"older decay exponents are allowed. It is, however, only ``almost optimal'' in terms of weighted Sobolev space decay: Indeed, the definition of ADM mass requires metrics in Sobolev spaces with a decay exponent in an interval $[1/2,1]$, and our Sobolev-space hypotheses exclude the  borderline case $1/2$. We plan to return to the missing borderline Sobolev case in the future.

Some comments on the proof are in order. As already pointed out, the first step is to establish existence of the retarded foliation $\mcN(u)$. There is a standard way of reducing this problem to a small-data existence result by scaling down, see the proof of Theorem~\ref{T9V16.1} below.
 One can then imagine adapting the existing global existence arguments to prove directly global existence for small data in domains of dependence, but this does not appear to be straightforward within the scheme of proof of~\cite{BieriJDG}, which is the only one so far under optimal conditions. Here we take a different route, by showing that scaled-down exterior regions can be filled-in by  initial data with small energy while preserving the vacuum constraint equations. This is done by a gluing argument which is relatively standard for initial data with well defined energy-momentum, center of mass, and angular momentum. This is the contents of Theorem~\ref{T9V16.1} below.
 {All proofs are given in detail so far.
 In Section~\ref{Structures} we outline the proof that the leaves $\mcN(u)$ have a well defined TB mass, and that the Trautman-Bondi mass loss holds. }

\section{Stability theorems}
 \label{s3V16.1}

In this section we review some results concerning vacuum stability   of Minkowski space-time, as needed for our purposes below.

We start with the following definition, from~\cite{BieriJDG} (see Appendix~\ref{s19V16.1} for notation):

\begin{Def} \label{intAFB}
An asymptotically Euclidean initial data set $({\R^3} ,g, K)$ is said to be an
\emph{AFB initial data set} if there exists a coordinate system $(x^1, x^2, x^3)$ outside of a ball such  that
\bea
    g_{ij} & = &  \delta_{ij}  +
    O_{H^3} (r^{- \frac{1}{2}}) \label{afgeng}
 \,,
\\
 K_{ij} & = &  O_{H^2} (r^{- \frac{3}{2}})
 \label{afgenk}
\end{eqnarray}
(see Appendix~\ref{s19V16.1} for notations).
\end{Def}
Set
\bea
\label{globalsalbQ}
 Q(a, x_{(0)})
  & =  &
   a^{-1} \int_{{\R^3}  _0}  \big(\
    |K |^2  +  (a^2 + d_0^2) |\nabla K |^2
\\
 \nonumber
 & &
 + (a^2 + d_0^2)^2 \ |\nabla^2 K |^2
\\ \nonumber
& &
  +
(a^2 + d_0^2)  |\mathrm{Ric}|^2
+ (a^2 + d_0^2)^2   |\nabla \mathrm{Ric}|^2   \big)   d \mu_{g}
  \, ,
\end{eqnarray}
where $a$ is a positive number related to the scale-covariance of the problem at hand,
$\nabla^\ell$ denotes the tensor of $\ell$-th covariant derivatives, and $d_0$ denotes the distance function from a chosen origin $x_{(0)}$.

Let $\inf_{x_{(0)}, a} Q (x_{(0)}, a)$ denote the infimum over all choices of origin $x_{(0)}$ and all
$a$ of the quantity defined by (\ref{globalsalbQ}).

A succinct version of the main result of~\cite{BieriJDG} can be formulated as follows:

\begin{The} \label{maintheoremlb2}
There exists $\epsilon>0$ such that for all smooth vacuum AFB initial data sets satisfying
\be
\inf_{x_{(0)}, a} Q (x_{(0)}, a)  \  <  \ \epsilon
\label{globalsalb2}
\end{equation}
the associated maximal globally hyperbolic vacuum development is geodesically complete, with the metric asymptotically approaching the Minkowski metric in all directions, and contains a \rfcp\   $
\mcN(u)$, $u\in \R$.
\end{The}

The above can be compared to the result of D. Christodoulou and S. Klainerman~\cite{Ch-Kl}, who consider the following class of initial data sets:

\begin{Def} \label{intSAFCK}
An initial data set $({\R^3} ,g,K)$ is called \emph{strongly asymptotically Euclidean}  if there exists a coordinate system
$(x^1, x^2, x^3)$ defined outside a compact set such that
\bea
{g}_{ij} \ & = & (1 +\frac{2m}{r}) \ \delta_{ij} +O_{H^4} (r^{- \frac{3}{2}})
 \,, \label{safg} \\
K_{ij} & = &   O_{H^3} (r^{- \frac{5}{2}})\,.  \label{safk}
\end{eqnarray}
\end{Def}

In order to state their global smallness assumption, for $b\in \R^+$ Christodoulou and Klainerman introduce
the
quantity
\bea
   \label{QforglobalsaCK1}
 \lefteqn{
Q_{CK}(x_{(0)}, b) =
\sup_{\R^3}    \big( \ b^{-2} (d_0^2 +b^2)^3  |\mathrm{Ric}|^2    \big)
}
 &&
\\
  & &  +
b^{-3}
  \int_{\R^3}\Big(    \sum_{l=0}^3 (d_0^2 +b^2)^{l+1}   |\nabla^l K |^2
 + \sum_{l=0}^1    (d_0^2 +b^2)^{l+3}
 |\nabla^l B |^2        \Big)
 \,,
  \nn
\end{eqnarray}
where  the
\emph{Bach tensor} $B$ is the following symmetric, traceless $2$-tensor field:
\[
B_{ij} \ = \ \epsilon_j{}^{k \ell}  \nabla_k ( R_{i\ell}  -  \frac{R}{4}   g_{i\ell}  ) \, .
\]
Christodoulou and Klainerman show that there exists $\epsilon>0$ such that the conclusions of Theorem~\ref{maintheoremlb2} hold for all vacuum initial data sets on $\R^3$ with $\tr_g K=0$ satisfying
\be \label{globalsmallnessassptck}
\inf_{x_{(0)} \in {\R^3}  , b \geq 0} \ Q_{CK}(x_{(0)}, b)  \ < \ \epsilon \,.
\end{equation}
Under these stronger conditions, they  show that the Hawking mass of a specific family of spheres has a limit when receding to infinity along the level sets of an outgoing null coordinate $u$. They call this limit the Bondi mass, though it should be said that the possibility of performing, under their hypotheses, the construction needed to define the mass  \emph{\`a la Bondi et al.} is not clear.%
\footnote{It follows e.g.\ from our discussion in Section~\ref{s6V16.5} that this limit will coincide with the Trautman-Bondi mass \emph{if} the construction of Bondi coordinates (compare Section~\ref{s6V16.1+}) can be carried out.}
They settle in the affirmative question \eq{28VIII16.1} for their definition of mass, but they do not prove that the past limit of their mass coincides with the ADM mass.
See Section~\ref{s2V16.2} for a further discussion.

The Christodoulou-Klainerman proof has been generalised to include Maxwell fields in~\cite{Zipser,BieriZipser2}.

Under the same restrictions on the fall-off on the initial data on a maximal slice but without any smallness conditions, Klainerman and Nicol\`o~\cite{KlainermanNicoloBook} prove that the maximal globally hyperbolic development of the initial data  contains a \rfcp.
In other words, the ``{future-complete null hypersurfaces theorem}'' holds under their hypotheses.  They also  settle in the affirmative questions \eq{28VIII16.1}-\eq{28VIII16.2} in their setting.

In~\cite{LindbladRodnianski2} global existence is proved for small perturbations of Minkowskian initial data of the form
\bea
{g}_{ij} \ & = & (1 +\frac{2m}{r}) \ \delta_{ij}  +  O_{H^7} (r^{- \alpha})
 \,, \label{safg-} \\
K_{ij} & = &   O_{H^6} (r^{- 1-\alpha})\,
,
 \label{safk-}
\end{eqnarray}
with some $\alpha >1$, using completely different methods. Those authors also allow a coupling to a massless scalar field. Here the smallness criterion is a weighted Sobolev norm as in \eq{safg-}-\eq{safk-}. Their argument has been generalised to include Maxwell fields and to all higher dimensions in~\cite{Loizelet:AFT}, compare~\cite{CCL}. Improved asymptotic estimates in the wave-coordinates setting of~\cite{LindbladRodnianski2} have been established in~\cite{LindbladAsymptotics}. These estimates allow one to answer positively \eq{28VIII16.1}-\eq{28VIII16.2} in the Lindblad-Rodnianski setting \cite{LindbladPrivate}.

\section{The mess with the mass}
 \label{s22X16/2}

Many authors use the term ``Bondi mass'' to denote rather distinct mass-like quantities defined in the radiation regime. This is very confusing, as more often than not it is not clear whether the object considered can be directly related to the mass as defined by Bondi et al. The aim of this section is to make a clear distinction between three notions of masses: the Trautman mass, the Bondi mass, and the mass of null hypersurfaces. While all three definitions give the same number whenever all three can be simultaneously defined, the prerequisites for each of the definitions are completely different. From this point of view the Bondi mass, as originally defined, appears to be the least general, since its existence implies the existence of the other ones.

\subsection{The Trautman mass}
 \label{s6V16.1}

In 1958 Trautman~\cite{Trautman58b,Tlectures} considers metrics satisfying
\beal{8V16.1}
 &
 g_{\mu\nu}-\eta_{\mu\nu} = O(r^{-1})
 \,,
 \quad
 \partial _\sigma g_{\mu\nu} = \hnews_{\mu\nu} k_\sigma + O(r^{-2})
 \,,
 &
\\
 &
 (\hnews_{\mu\nu} - \frac 12 \eta^{\alpha \beta} \hnews_{\alpha \beta} \eta_{\mu \nu}) \eta^{\nu\rho} k_\rho = O(r^{-2})
 \,,
 &
\eeal{8V16.2}
with some tensor $N_{\mu\nu}$, where $\eta_{\alpha \beta}$ is the Minkowski metric and $k_\mu$ is a null vector field which will be assumed here to asymptote to
\bel{8VI16.3}
 k_\mu dx^\mu \to_{r\to\infty} dt - dr
 \,.
\end{equation}
Equations~\eq{8V16.1} and \eq{8VI16.3} are motivated by the behaviour of solutions of  wave equations at large retarded times~\cite{Trautman58b}, while \eq{8V16.2} is a precise version of the ``asymptotically harmonic coordinates'' condition.

Given a two-surface $S$ one sets%
\footnote{See~\cite{Trautman58b} for the original motivation of the definition. A Hamiltonian analysis leading to \eq{toto_new}-\eqref{Freud2.0_new} can be found in~\cite{CJK}, compare~\cite{ChAIHP}.}
\begin{eqnarray}
  H(X,S)&= &\frac 12 \int_{S}
 \ourU^{\alpha\beta}dS_{\alpha\beta}\,,
\label{toto_new}
\end{eqnarray}
where is any Killing vector of the Minkowski metric $\bMinkowskimetric$. Here $d
S_{\alpha\beta}$ is defined as
$\frac{\partial}{\partial x^\alpha}\lrcorner \frac{\partial}{\partial
  x^\beta}\lrcorner \rd x^0 \wedge\cdots \wedge\rd x^{n} $, with
$\lrcorner$ denoting contraction, and $\ourU^{\alpha\beta}$ is given by
\begin{eqnarray}
  \ourU^{\nu\lambda}&:= &
{\ourU^{\nu\lambda}}_{\beta}X^\beta
  + \frac 1{8\pi} \Delta^{\alpha[\nu}
  {X^{\lambda]}}_{;\alpha}
\ ,
 \phantom{xxx}
 \label{Fsup2new_new}
\\ {\ourU^{\nu\lambda}}_\beta &:= & \displaystyle{\frac{2|\det
  \bMinkowskimetric _{\mu\nu}|}{ 16\pi\sqrt{|\det g_{\rho\sigma}|}}}
g_{\beta\gamma}(e^2 g^{\gamma[\nu}g^{\lambda]\kappa})_{;\kappa}
\,,\label{Freud2.0_new}
\end{eqnarray}
where a
semicolon  denotes covariant
differentiation \emph{with respect to the Minkowski metric $\bMinkowskimetric$},
while
\bea
  \label{mas2_new}
   &
    e := \frac{\sqrt{|\det g_{\rho\sigma}|}}{\sqrt{|\det\bMinkowskimetric _{\mu\nu}|}}
     \;,
     \qquad
 \Delta^{{  \alpha}\nu}:=e\, g^{{ \alpha}\nu}- \bMinkowskimetric{}^{{ \alpha}\nu}
    \,.
   &
\end{eqnarray}

Let $S_{t,r}$ denote a surface of constant $t$ and $r$ and let $u=t+f(\vec x)$ satisfy
$\partial_\mu u = k_\mu$.
Trautman argues that the limit%
\footnote{Strictly speaking, Trautman does not say explicitly that the limit in \eq{8VI16.3} should be taken at fixed $u$, but this is suggested by the discussion in the accompanying paper~\cite{Tlectures}. Further, he only notes that the potentially divergent terms in the integrand cancel out, without justifying the existence of the limit. The latter is clear if one assumes that $\hnews_{\mu\nu}$ is of the form $f_{\mu\nu}(u,\theta,\varphi)/r$ for some continuous functions $f_{\mu\nu}$, as again suggested by the context in~\cite{Trautman58b,Tlectures}. We also note that the term containing derivatives of $X$ in \eqref{Fsup2new_new} does not appear in~\cite{Trautman58b} since only translations are considered there.}
\bel{8V16.3+}
 p^T_\mu(u) := \lim_{r\to\infty} H(\partial_\mu,S_{u+f(\vec x),r})
\end{equation}
exists,  and satisfies the \emph{Trautman mass-loss formula}
\bea
 \label{8V16.4}
 \lefteqn{
 p^T_\gamma(u_2) - p^T_\gamma(u_1)=
  \frac {1} {4\pi}
  \int_{u=u_1}^{u_2}\int_{S^2} \eta^{\alpha\rho}\eta^{\beta\sigma}\hnews_{\rho\sigma}(\hnews_{\alpha\beta}
 }
 &&
\\
 \nn
 &&
  - \frac 12 \eta^{\mu\nu} \hnews_{\mu\nu}\eta_{\alpha\beta}) k_\gamma \, du \, \sin(\theta) \, d\theta\, d\varphi
  \,,
   \phantom{xxx}
\eea
where the integrand is non-negative by \eq{8V16.2}.

We will refer to $p_\mu^T$ as the \emph{Trautman four-momentum}, and to $p_0^T$ as the \emph{Trautman mass}.

Let $T_{R,u}$ denote the timelike cylinder
\bel{8V16.5}
 T_{R,u}=\{|\vec x|=R,\ 0\le t \le u+f(\vec x)
 \}
 \,,
\end{equation}
where we have assumed that $u$ is a null coordinate of the form $u=t-f(\vec x)$. The divergence identity on $T_{R,u}$ gives
\bea
H(\partial_\mu, S_{u+f,R}) - H(\partial_\mu, S_{0,R})
  & = &
   \int_{T(R,u)}
 \partial_\alpha \ourU^{\alpha\beta}dS_{\beta}\,.
\end{eqnarray}
It is well known that the limit, as $r$ tends to infinity, of the integrals $H(\partial_\mu, S_{t,r})$ at \emph{fixed $t$} is the ADM four-momentum $p_\mu$ under the decay conditions \eqref{afgeng}-\eqref{afgenk} on the initial data. Passing to the limit $R\to\infty$ we find
\bel{8V16.6+}
 p^T_\mu(u) - p_\mu
 = \lim_{R\to\infty} \int_{T(R,u)}
 \partial_\alpha \ourU^{\alpha\beta}dS_{\beta}\,.
\end{equation}
We conclude that:

\begin{Proposition}
  \label{P8V16.7}
The past limit $\lim_{u\to-\infty} p^T_\mu(u)$ of the Trautman four-momentum equals the ADM four-momentum if and only if
\bel{8V16.6}
 \lim_{u\to-\infty}
 \Big(\lim_{R\to\infty} \int_{T(R,u)}
 \partial_\alpha \ourU^{\alpha\beta}dS_{\beta}
  \Big)
 =0
 \,.
\end{equation}
\end{Proposition}

\subsection{The Trautman-Bondi mass}
 \label{s6V16.1+}

The definition of mass  introduced in 1962 in~\cite{BBM,Sachs}  requires existence of a coordinate system in which the metric takes the form
\bel{19II16.1}
	 \og = \og_{00} du^2 -2 e^{2\beta } dr \,du - 2 r^2 U_A dx^A du + r^2  \underbrace{h_{AB}  dx^A dx^B}_{=:h}
	\,,
\end{equation}
where the determinant of $h_{AB}$ is $r$-independent.
The authors of~\cite{BBM,Sachs} further require  the fields $g_{00}, \, U_A, \, \beta $ and $h_{AB}$ to have full asymptotic expansions in terms of inverse powers of $r$, with
\bea
 &
 h_{AB} = \mathring h_{AB} + O(r^{-1})
  \,,
  \quad
  \beta = O(r^{-2})
  \,,
  \quad
  U_A = O(r^{-2})
  \,,
&
\\
&
 \displaystyle
  g_{uu}  = -1 + \frac{2\mu_B(u,x^A)} r + O(r^{-2})
 \,.
 &
\eeal{6VI16.10}
The function $\mu_B$ is called the \emph{Bondi mass-aspect function}.

Under the conditions spelled-out above, one can introduce ``asymptotically Minkowskian'' coordinates $x^\mu$ by setting
\bel{6V16.13}
 t = u + r
 \,,\
 x = r \sin(\theta)\cos(\varphi)
 \,,\
 y = r \sin(\theta)\sin(\varphi)
 \,,\
 z = r \cos(\theta)
 \,.
\end{equation}
Here $\theta$ and $\varphi$ are coordinates on $S^2$ in which
$$
 \mathring h_{AB}dx^Adx^B = d\theta^2 + \sin^2(\theta) d\varphi^2
  \,.
$$
A calculation shows that the asymptotic conditions
\eq{8V16.1}-\eq{8VI16.3} are satisfied, and that the Trautman mass $p^T_0$ defined in \eq{8V16.3+} equals the right-hand side of the formula
\bel{eq:mTB}
	\mTB(u) := \frac{1}{4\pi} \int_{S^2} \mu_B(u,\theta,\varphi) \, \sin(\theta)\, d\theta\,d\varphi
 \,,
\end{equation}
first proposed in~\cite{BBM,Sachs}. From this point of view, Trautman's definition is thus more general that of Bondi et al, and precedes the latter.
We will refer to $\mTB$ as the \emph{Trautman-Bondi mass}.

Existence of  coordinates \eq{19II16.1}-\eq{6VI16.10} in asymptotically vacuum space-times   admitting smooth conformal completions has been established
in~\cite{TamburinoWinicour} when $\Lambda=0$, and in~\cite{ChMS} for polyhomogeneous $\Scri$'s; this is easily generalised to $\Lambda\in\R$, see~\cite{ChIfsits}.

\subsection{The mass of characteristic hypersurfaces}
 \label{s6V16.5}

Now, one does \emph{not} expect to have expansions in powers of $r^{-1}$ in general physically relevant vacuum space-times to sufficiently high order to be able to carry out the construction needed for the definition of the Bondi mass; in particular, no such expansions are expected in general small-data spacetimes of~\cite{Ch-Kl,KlainermanNicoloBook,BieriJDG,LindbladRodnianski2} discussed in Section~\ref{s3V16.1}. An alternative construction of a mass-type object can be carried-out when considering null hypersurfaces, without the need to introduce Bondi coordinates and high-order expansions. Since the resulting mass coincides with the Trautman-Bondi mass in situations where Bondi's construction can be carried out as well, it has become standard to retain the name of Trautman-Bondi mass in this context.

Consider, then, a null hypersurface $\mcN$.  Near $\mcN$ we can introduce a coordinate system $(x^0\equiv u, x^1 \equiv r, x^A)$ so that $\mcN=\{u=0\}$, $  r$ is a coordinate parameterising the null geodesics threading $\mcN$, and $x^A$ are coordinates parameterising those geodesics.  Putting an overbar to denote restriction of a field to $\mcN$, the spacetime metric at $u=0$ takes the form
\begin{equation}
\overline{g}:=g|_{x^{0}=0}\equiv
 \overline{g}_{00}(dx^{0})^{2}
+2\overline{g}_{01}dx^{0}dx^{1}
+2\overline{g}_{0A}dx^{0}dx^{A}
+\overline{g}_{AB}dx^{A}dx^{B}
\;.
\label{null2}
\end{equation}
Let us denote by $\theta^+$ the divergence of the generators of $\mcN$,
\bel{6V16.21}
 \theta^+ := \frac 12 \ol g^{AB} \partial_r \ol g_{AB}
 \,.
\end{equation}
Now,  each section
$$
 S_\rho:=\{r=\rho\} \subset \mcN
$$
of $\mcN$ defines, at least locally,  two null hypersurfaces normal to  $\mcN_\rho^\pm$, one of them being $\mcN$, say $\mcN = \mcN_\rho^+$. These hypersurfaces are obtained by shooting null geodesics in null directions orthogonal to $S_\rho$. Let $\ell^-$ denote a field of null tangents to $\mcN_\rho^-$, we can normalise $\ell^-$    by requiring that
$$
 g(\partial_r,\ell^-) = -2
 \,.
$$
We will denote by $\theta^-$ the divergence of $\mcN_\rho^-$ with respect to this normalisation.

Suppose that $\mcN_u$ is one of the level sets of the Bondi coordinate $u$ as described at the beginning of Section~\ref{s6V16.1+}. One then has the expansions%
\footnote{See, e.g.,~\cite{ChPaetzBondi}. In that reference $\theta^+$ is denoted by $\tau$, and $\theta^-$ by $\zeta$.}
\bel{6VI1.6.21}
 \theta^\pm = \pm \frac 2 r + \frac{\theta^\pm_2 (x^A)}{r^2} + o(r^{-2})
 \,,
\end{equation}
for some functions $\theta^\pm_2$ on $S^2$.
It has been shown in~\cite{ChPaetzBondi} that
\bel{6VI16.22}
 \mTB = \frac 1 {4\pi} \int_{S^2} (\theta^+ + \theta^-) \, \sin(\theta)\, d\theta \, d\varphi .
\end{equation}
As explained in~\cite{ChPaetzBondi}, the advantage of \eq{6VI16.22} as opposed to \eqref{eq:mTB}, is that $\theta^\pm$ can be defined purely in terms of the characteristic initial data on $\mcN$, without imposing any coordinate conditions away from $\mcN$. Furthermore, the coordinate $r$ does not have to be a Bondi radial coordinate, because the integrand of \eq{6VI16.22} is
invariant under redefinitions of $r$ which preserve the asymptotic behaviour of the intrinsic tensor field $g_{AB}dx^A dx^B$~\cite{ChPaetzBondi}.

For those reasons it appears natural to view the right-hand side of \eq{6VI16.22} as a natural  {definition} of the mass of characteristic hypersurfaces extending to infinity in asymptotically Minkowskian space-times.

It follows immediately from the asymptotic expansions \eq{6VI1.6.21}, together with%
\footnote{We are grateful to Tim Paetz for pointing this out.}
\begin{eqnarray}
 \label{6VI16.23}
\sqrt{\mathrm{det} \ol g_{AB}}
  &=&
   r^2\sqrt{\mathrm{det} \mathring h_{AB}} \Big(1-\tau_2r^{-1} + O(r^{-2})\Big)
   \,,
\end{eqnarray}
that
\bel{6VI16.22+}
 \frac 1 {4\pi}  \int_{S^2} (\theta^+ + \theta^-) \, \sin(\theta)\, d\theta \, d\varphi
 =
 \lim_{r\to\infty} m_H(S_r)
 \,.
\end{equation}
Here $m_H$ is the \emph{Hawking mass} of the spheres of constant $r$ within $\mcN$:
\begin{equation}
 m_{\mathrm{H}}(S_r) = \sqrt{\frac{\mathrm{Area}(S_r)}{16\pi}}\Big(1+\frac{1}{16\pi}\int_{S_r}\theta^+ \theta^-\, d\mu_{\ol g}
  \Big)
\;,
\end{equation}
with $d\mu_{\ol g}=\sqrt{\det \ol g_{AB}}\, d^2 x $ being the measure induced by $\ol g$ on $S_r$. Thus, the mass of characteristic hypersurfaces extending to $\scrip$ is the limit of the Hawking masses of suitable coordinate spheres within the hypersurface.

\subsection{A summary}
 \label{ss25XI16.1}

Our discussion so far can be summarised as follows:

\begin{enumerate}
 \item Historically, the first definition of mass at null infinity has been given by Trautman, who also proved its monotonicity~\cite{Trautman58b}.
  \item Trautman's definition applies to situations more general than the one considered by Bondi et al., and reduces to the Bondi definition when Bondi coordinates exist.
  \item The Hawking mass of suitable spheres has a limit, when receding to infinity  along outgoing null hypersurfaces, in all the settings where it has been analysed so far: the Christodoulou-Klainerman proof~\cite{Ch-Kl} and its generalisation by Bieri~\cite{BieriJDG} the Klainerman-Nicol\`o analysis~\cite{KlainermanNicoloBook}, and the  characteristic initial-data approach of~\cite{ChPaetzBondi}.
      The limit, which will be referred to as the \emph{Trautman-Bondi mass}, coincides with the Bondi mass whenever Bondi coordinates exist as well.
      \item The existence of Bondi coordinates has only been established under very restrictive hypotheses so far: for initial data stationary outside of a compact set, or for the conformally-smooth hyperboloidal or characteristic Cauchy problem, or assuming a smooth or polyhomogeneous $\scri$.
          \item The notion of the mass at null infinity within the framework of Lindblad-Rodnianski~\cite{LindbladRodnianski2} remains to be clarified.
\end{enumerate}

\section{``Poincar\'e charges''}
 \label{s6V16.21}

In this section we review the conditions which need to be satisfied to be able to define the total  energy-momentum, angular momentum and the center of mass of  asymptotically Euclidean initial data sets.

Let $\alpha\in\R^+$, $\ell \in \N$, $\ell\ge 1$.
We shall say that $(g,K)$ on the exterior $E$ of a ball in
$\mathbb{R}^3$ constitutes  an $C^{-\alpha}_\ell$-\emph{asymptotically Euclidean end}
 provided there are coordinates in which, for all
multi-indices $|\gamma|\leq \ell$, $|\beta|\leq \ell-1$,
\begin{equation} \label{af}
|\partial^{\gamma}(g_{ij}-\delta_{ij})(\vec   x)|=O(|\vec   x|^{-|\gamma|-\alpha})
 , \qquad
 |\partial^{\beta}K_{ij}(\vec   x)|=O(|\vec   x|^{-|\beta|-1-\alpha}),
\end{equation}
where $\partial$ denotes the partial derivative operator.  Note that the index $\ell$ refers to the
differentiability class of the metric, with $K$ being a priori only $(\ell-1)$-times differentiable.
Throughout the rest of this work we require  $\ell
\geq 3$, compare Remark~\ref{R1V16.2} below.
We say that $(M,g,K)$ is $C^{-\alpha}_\ell$-\emph{asymptotically
Euclidean} (AE) if $M$ is the union of a compact set and a finite
number of ends, all of  which are $C^{-\alpha}_\ell$-asymptotically Euclidean.

An obvious analogue of the above are the definitions of $W^{-\alpha}_{\ell,q}$-asymptotically Euclidean manifolds and ends, where one requires
that in each end we have
\bel{20V16.4}
 g-\delta \in W^{-\alpha}_{\ell,q}
 \ \mbox{ and } \ K \in  W^{-\alpha-1}_{\ell-1,q}
 \,.
\end{equation}

As pointed out
in~\cite{ChErice},
every $C^{-\alpha}_\ell$-asymptotically Euclidean end  with $\alpha>1/2$ possesses a
well-defined finite energy-momentum vector $(p_0,\vec p)$ when the dominant energy condition $|\vec J| \le \rho$ holds with $\rho \in L^1$. This remains true for $W^{-\alpha}_{\ell,q}$-asymptotically Euclidean ends with $\alpha \ge 1/2$ and $q \ell>3$.

Further conditions have to be imposed on the initial data to guarantee convergence of the integrals defining the centre of mass and total angular momentum.
One possible such condition is a \emph{parity requirement} (cf.~\cite[Proposition~E.1]{ChDelay}), that there exists $\alpha_- >0$ satisfying %
\bel{15VIII16.1}
 \alpha+\alpha_->2
\end{equation}
such that  we have
\bea
&
 \big| g_{ij}(\vec   x)-g_{ij}(-\vec   x) \big|
 +(1+r)\big|\partial_k \big(g_{ij}(\vec   x)-g_{ij}(-\vec   x)\big)\big|
  =O(|\vec   x|^{ -\alpha_-})
 \,,
 &
\\
 &
 \big | K_{ij}(\vec   x)+K_{ij}(-\vec   x) \big|=O(|\vec   x|^{-1- \alpha_-})
  \,.
  &
  \label{29IV16.2+}
\end{eqnarray}
This requirement is in the spirit of, but weaker than the Regge-Teitelboim conditions for a well-defined angular-momentum and center of mass, which are \eq{29IV16.2+} with $\alpha=1$ and $\alpha_-=2$.

The parity conditions above are satisfied by the initial data sets of Christodoulou-Klainerman, Klainerman-Nicol\`o and Lindblad-Rodnianski discussed in Section~\ref{s3V16.1}; they are not necessarily so by the AFB initial data of~\cite{BieriJDG}.

Compare~\cite{Changmom} for further classes of initial data sets with a well defined angular momentum.

\section{Interior gluing}
 \label{s29IV16}

The aim of this section is to show how to fill-in vacuum initial data sets defined in an exterior region to initial data on $\R^3$. This will be done so that small exterior data will have a small interior filling.

{It is shown in~\cite{ChBoost} how to carry out such a construction for initial data with
 a well-defined total energy-momentum vector.}
Here we give a simpler proof  of existence of fillings satisfying \eq{globalsalb2} under the supplementary hypothesis of a well defined center of mass and total angular momentum.

 Let
$$
Q:=(p_0,\vec p,  \vec c, \vec J\,)
$$
be the ``Poincar\'e charge'' of  suitably asymptotically Euclidean initial data, as discussed in
Section~\ref{s6V16.21}.
 Here  $p_0$ is the total ADM energy, $\vec p$ the total ADM momentum, $\vec c$ the center of mass and $\vec J$ the total ADM angular-momentum. As a first step, we show that every $Q$ can be realised by an initial data set $(\hyp,g_Q,K_Q)$ with well controlled asymptotics, and such that the norms
of  $(g_Q,K_Q)$ relevant for global existence theorems tend to zero when $Q$ tends to zero in a way compatible with scaling-down a vacuum initial data set with well defined $Q$.

We note that our definition of $\vec c$ and $\vec J$ differs from that in~\cite{HSW}, in that  their $(\vec c,\vec J)$ equals our $(\vec c,\vec J)$ divided by $p_0$. This only introduces trivial notational changes in the arguments below (affecting e.g.\  the scaling behaviour of $\vec c$ and $\vec J$).

\medskip

We have:

\begin{Proposition}
 \label{P1V16.1}
 \begin{enumerate}
   \item
    \label{p1V16.1}
For any Poincar\'e charge $Q$ with timelike vector $(p_0,\vec p)$ there exists a smooth vacuum initial data set $(\R^3, g,K)$ satisfying the Regge-Teitelboim parity conditions \eqref{29IV16.2+} with $\alpha=1$, $\alpha_-=2$, and realizing $Q$.
 \item
    \label{p1V16.2}
    Let $0<\eta<1$ and let
    \bel{5V16.1}
      \mbox{$\theta \subset \R^{10}$ be a compact set}
    \end{equation}
     of charges $Q\equiv (p_0,\vec p, \vec c, \vec J)\equiv (p , \vec c, \vec J)$ such that for all $Q\in \theta$ we have $|\vec p\normwithnodelta   \le \eta p_0$, where $|\vec p\normwithnodelta  $ is the Euclidean norm of $\vec p$. Consider
    a one-parameter family of Poincar\'e charges of the form
    \bel{5V16.2}
     Q(\epsilon)=\epsilon \big(\hat p(\epsilon),
     \epsilon \vec{\hat  c}(\epsilon),
      \epsilon \vec{\hat  J}(\epsilon)\big)
     \,,
     \quad (\hat p(\epsilon),    \vec{\hat  c}(\epsilon),
      \vec{\hat  J}(\epsilon)\big) \in \theta\,,
    \end{equation}
    with $\epsilon$ belonging to a neighborhood of zero in $\R$.
    Then for each  Poincar\'e charge  $Q(\epsilon)$ as in \eq{5V16.2} there exists an initial data set  $(\R^3, g(\epsilon),K(\epsilon))$ satisfying the parity conditions  \eqref{29IV16.2+} with $\alpha=1$, $\alpha_-=2$ and realizing $Q(\epsilon)$ so that  the norm $C^{-1}_{k+1}\oplus C^{-2}_{k}$ of $( g(\epsilon),K(\epsilon))$ tends to zero as $\epsilon$ goes to zero for any $k\in\N$.
 \end{enumerate}
\end{Proposition}

\begin{Remark}
  \label{9V16.1}
 {\rm For the initial data of point~\ref{p1V16.2}  the infimum $\inf_{x_{(0)}, a} Q (x_{(0)}, a)$ of  (\ref{globalsalbQ})  over all choices of origin $x_{(0)}$ and all
$a$ tends to zero as $\epsilon$ goes to zero.
}
\qed
\end{Remark}

\begin{Remark}
  \label{9V16.1+}
 {\rm One can use~\cite[Lemma~5.2]{HSW} and~\citep{CCI2} to arrange  the above data sets  to coincide with e.g.\ Kerr outside of a sufficiently large compact set, but this is irrelevant for our purposes.
}
\qed
\end{Remark}

\begin{proof}
\ref{p1V16.1}. The result is the contents of~\cite[Corollary~5.3]{HSW}, we give our version of its proof as a warm-up to the proof of part \ref{p1V16.2}. Let $g$ be any smooth metric on $S^3$ with strictly positive scalar curvature which is
\emph{not} conformally flat. While this is not necessary, for definiteness we also require that the metric be the standard unit round sphere on some open subset of $S^3$. Let $\varphi$ be the Green function of the operator $\Delta - R/8$, where $R$ is the scalar curvature of $g$. Again for definiteness, we take the singularity to be located at an interior point of the region where the metric is round. Then $\phi^4 g$ is a smooth AE manifold with zero scalar curvature.

We will refer to the metric induced on the usual static slices of the Schwarzschild space-time as the space-Schwarzschild metric.
By~\cite{Corvino} we can deform $\phi^4 g$ to a zero-scalar-curvature metric $\hat g$ which coincides with the space-Schwarzschild metric outside of a compact set. The metric $\hat g$ has strictly positive energy $p_0$ by the positive mass theorem.

Let $(\mcM,\fourg)$ be the maximal globally hyperbolic vacuum development of $(\R^3, \hat g,K)$ with $K\equiv 0$. Then the metric $\fourg$ coincides with the Schwarzschild metric in the domain of dependence of the region where $\hat g$ is the space Schwarzschild metric. In particular $(\mcM,\fourg)$ contains AE spacelike hypersurfaces with arbitrary timelike four-momentum $p^\mu$, with $(p_0)^2-|\vec p|^2 = m^2$, obtained by asymptotically performing an active Lorentz boost to the initial data surface. The initial data on the boosted slices will satisfy the original Regge-Teitelboim parity conditions, with $\alpha=1$ and $\alpha_-=2$, see Appendix~\ref{A3VI16.1}.

Let $\lambda>0$, then the metric $\lambda^2 \hat g$ is AE and has mass $m \lambda$. To see this, let $(x^i)$ be a global coordinate system in which $\hat g$ takes the form
$$
 \hat g =  \hat g_{ij}(\vec x) dx^i dx^j = \bigg(
  \Big(1+\frac{2m}{|\vec x|}\Big)\delta_{ij} + O(1/|\vec x|^2)\bigg) dx^i dx^j
 \,.
$$
In rescaled  coordinates $\vec y =\lambda \vec x $ we have
$$
   \lambda^2 \hat g =  \hat g_{ij}(\lambda^{-1} \vec y) d(\lambda x^i) d(\lambda x^j)
   =\bigg(
   \Big(1+\frac{2 \lambda m}{|\vec y|}\Big)\delta_{ij}
     + \lambda^2 O(1/|\vec y|^2) \bigg) dy^i dy^j
 \,,
$$
as desired. As $\lambda$ varies over $\R^+$, the collection of  maximal globally hyperbolic developments of the associated time-symmetric initial data sets contains boosted slices with any timelike energy-momentum vector.

By~\cite[Theorem~3]{HSW}, each of the above  initial data sets can be deformed to a new initial data set with the same energy-momentum vector $p^\mu$,  arbitrary angular-momentum, and arbitrary center of mass.

\medskip
\ref{p1V16.2}: We start by noting that, in view of point \ref{p1V16.1}, it suffices to construct the desired initial data sets for sufficiently small $\epsilon$.

Let $\varepsilon >0$, and let $g(\varepsilon)$ be a smooth family of smooth metrics on $S^3$ which coincides with the unit round metric, say $g_0$, on an open neighborhood of the south pole of $S^3$, with $g(\varepsilon)$ tending to $g_0$ as $\varepsilon$ approaches zero in $C^\infty(S^3)$. We further assume that all the metrics $g(\varepsilon)$ are symmetric with respect to reflection of $(S^3,g_0)$ across the north pole (by this we mean the map which to a point lying a distance
$0\le s\le \pi$ from the north pole on a $g_0$-geodesic starting at the north pole with tangent $\ell$ assigns the point lying the distance $s$ on the geodesic starting at the north pole with tangent $-\ell$), and none of them is  conformally flat; the existence of many such families of metrics  follows
from~\cite{ChBeignokids}.
 Conformally rescaling by a Green function $\varphi(\varepsilon)$ with the singularity at the south pole as in the proof of point~\ref{p1V16.1},
we obtain a family $\varphi^4(\varepsilon) g(\varepsilon)$ of AE metrics on $\R^3$ which tend to the flat metric as $\varepsilon$ tends to zero in $C_k^{-1}(\R^3)$ for any $k\in \N$, and which are invariant under the parity map $\vec x\mapsto - \vec x$.
Since the ADM mass is continuous in this topology, we obtain a family of metrics with ADM masses $m(\varepsilon)$ covering a neighborhood of zero, with vanishing momentum, angular momentum, and center of mass.

Let $(\mcM,\fourg(\varepsilon))$ denote the maximal globally hyperbolic vacuum development of $(\R^3,\varphi^4(\varepsilon) g(\varepsilon),K)$, with $K\equiv 0$.
By the Boost Theorem~\cite{christodoulou:murchadha} the spacetime $(\mcM,\fourg(\varepsilon))$ contains complete AE boosted hypersurfaces, the ADM four-momentum  $p(\varepsilon):=(p_0 (\varepsilon),\vec p (\varepsilon))$ of which takes arbitrary values satisfying $(p_0 (\varepsilon))^2 -|\vec p (\varepsilon)\normwithnodelta  ^2 = m^2 (\varepsilon)$.
The usual transformation law of global charges~\cite[Proposition~E.1]{ChDelay} shows that the resulting  initial data sets, which we denote by
 \bel{5V16.3}
  \Big(\R^3,g\big(p(\varepsilon)\big),K\big(p(\varepsilon)\big)\Big)
  \,,
 \end{equation}
have vanishing center of mass and vanishing total angular momentum. The metrics $g(p(\varepsilon))$ can be chosen to be parity-symmetric and the extrinsic curvature tensors $K(\varepsilon))$ can be chosen to be parity-antisymmetric.

Let $\hat \eta \in \R$ satisfy $\eta<\hat \eta <1$.
Let $\hat \theta$ be a compact neighborhood of the set  $\theta$ of \eqref{5V16.1} such that $|\vec p\normwithnodelta   \le \hat \eta p_0$ for all $(p_0,\vec p, \vec c, \vec J)\in \hat \theta$.
Let $\hat \theta_{p}$ denote the compact set of ADM four-momenta
$p:=(p_0,\vec p)$ obtained by projecting the set $\hat \theta$  on the four-momentum factor of $\R^{10}$.
Restricting oneself to boosted slices with $p(\varepsilon)  \in  \epsilon \hat \theta_p$, by the continuous-dependence-upon-initial-data results of~\cite{christodoulou:boost} one obtains a family of initial data sets \eq{5V16.3} such that $(\R^3,g(p(\varepsilon)),K(\varepsilon))$ tends to $(\R^3,\delta,0)$ in $C^{-1}_k\oplus C^{-2}_k$
for any $k\in \N$ as either $\epsilon$ or $\varepsilon$ tends to zero.

Let $\varepsilon_1$ be such that
\bel{5V16.5}
  \big( Q\in \theta \,, \ |Q-Q'\normwithnodelta   < 3 \varepsilon_1 \big)
 \
 \Longrightarrow
 \
 Q'\in \hat \theta
 \,.
\end{equation}
where $|Q-Q'\normwithnodelta   $ is the Euclidean distance in $\R^{10}$. In other words, a $3\varepsilon_1$-thickening of $\theta$ is included in $\hat \theta$.
Let us write $Q_{p(\varepsilon)}$ for those $Q\in \epsilon \hat \theta$ for which  $p=p(\varepsilon)$; equivalently, which are of the form $(p(\varepsilon),\vec c,\vec J)$.
For every such $Q$ we can carry-out the construction in~\cite{HSW} to obtain a vacuum initial data set, which we denote by
 \bel{5V16.4}
  \Big(\R^3,g\big(Q_{p(\varepsilon)}\big),K\big(Q_{p(\varepsilon)}\big)\Big)
  \,,
 \end{equation}
with the following properties: if we denote by $Q'_{{p(\varepsilon)}}$ the Poincar\'e charge of the data set \eq{5V16.4}, then a)
\bel{5V16.7}
 |Q'_{{p(\varepsilon)}}-Q_{p(\varepsilon)} | \le \varepsilon_1
 \,,
\end{equation}
and b) the projection of $Q'_{{p(\varepsilon)}}$ on the four-momentum factor of $R^{10}$ is ${{p(\varepsilon)}}$.
Here one should keep in mind that the construction in~\cite{HSW} is done in two steps: In the first step one adds a correction to the initial data at a large distance to approximately realize the charge $ {Q_{p(\varepsilon)}}$. In the second step, based on~\cite{CorvinoSchoen2}, one solves a suitable PDE  to ensure that the vacuum equations are satisfied. The distance from the origin to the region where the perturbation of the first step is localised depends upon $ {Q_{p(\varepsilon)}}$,
but can be chosen to be bounded independently of $ {Q_{p(\varepsilon)}}$ in our context because of compactness of $\hat \theta$. This implies that, for all $k\in \N$, the $C^{-1}_k\oplus C^{-2}_k$ norm of the first-step correction goes to zero as $\varepsilon$ goes to zero.
The fact that the $C^{-1}_k\oplus C^{-2}_k$ norm  of the second-step-correction goes to zero as well when $\varepsilon$ tends to zero is then a routine property of elliptic equations in weighted spaces (cf., e.g.,~\cite{ChoquetBruhatChristodoulou81,Bartnik86,ChAFT}).

It remains to show that the collection of Poincar\'e charges
$
  Q'_{{p(\varepsilon)}}
$
so obtained covers $\epsilon \theta$ for all $\epsilon$ small enough. This  follows from~\cite[Lemma~5.2]{HSW} using \eqref{5V16.5} and \eq{5V16.7}.
\qed
\end{proof}

We can use the solutions above to fill-in vacuum initial data sets defined on \emph{exterior regions}:

\begin{proposition}
 \label{Pkerr}
  Let $\N\ni\ell\ge 4$,  $\varepsilon>0$, $q>1$ and  $\alpha\ge 1/2$.
Let $(g,K)$ be $W^{-\alpha}_{\ell,q}$-AE vacuum initial data set (as defined at the beginning of Section~\ref{s6V16.21})
on the exterior of a
ball in $\mathbb R^3$, which we denote by  $E$,
with time-like ADM four-momentum
$(p_0,\vec p)$.
For sufficiently large $R$,
there is a vacuum initial data set $(\bar g, \bar K)$ on $E$ so
that on $E\cap \{ |\vec   x|\geq 2R\}$ we have  $(\bar g, \bar K)=(g,K)$, while on $E\cap\{
|\vec   x| \le R\}$ the initial data set $(\bar g, \bar K)$ is identical to one of the initial data sets of Proposition~\ref{P1V16.1}, point~\ref{p1V16.1}.
If $( p_0 + \delta p_0, \vec  { p}+\delta \vec p)$ is the four-momentum
of this last data set, then
\bel{2V16.1}
\mbox{$| \delta p_0|  < \varepsilon$ and
$\left | \delta \vec p\right| <\varepsilon.$}
\end{equation}
If, moreover, \eq{15VIII16.1}-\eq{29IV16.2+} hold, so that $(g,K)$ has a well-defined center of mass $\vec c$ and total angular momentum $\vec J$,
then the interior filling can be chosen as in point~\ref{p1V16.2} of Proposition~\ref{P1V16.1}, with
\bel{2V16.2}
 \mbox{$|\delta \vec   c|<\varepsilon$ and
 $|\delta \vec   J|<\varepsilon$.
 }
\end{equation}
\end{proposition}

\begin{Remark}
  \label{R2V16.1}
  {\rm
  The interior data sets are taken to be those of Proposition~\ref{P1V16.1} for definiteness, keeping in mind that we seek an interior solution with small relevant norms when $(E,g,K)$ has small norm. One can construct interior initial data set such that \eq{2V16.1} holds by choosing as the family of interior solutions \emph{any} family of initial data sets smoothly parameterized by ADM energy-momentum vectors belonging to a neighborhood of $(p_0,\vec p)$.  To obtain \eq{2V16.1}-\eq{2V16.2} one can take as interior solutions \emph{any} family of initial data sets smoothly parameterized by Poincar\'e charges belonging to a neighborhood of the Poincar\'e charge $Q$ of $(E,g,K)$.

While in this work we aim at a family of initial data on $\R^3$,  it suffices moreover that the ``interior initial data'' are defined on an annulus. In particular the gluing can be done so that the interior metrics are members of the Kerr family.
   }
\qed
\end{Remark}

\begin{Remark}
 \label{R1V16.1}
{\rm
Equations~\eqref{afjest}-\eqref{afjest0} below show that the hypothesis that the decay rate $\alpha$ is equal to one, made in~\cite[Proposition~3.2]{CCI2}, can be replaced by $\alpha>1/2$ in weighted H\"older spaces, or $\alpha\ge 1/2$ if using weighted Sobolev spaces.
%
}
\qed
\end{Remark}

\begin{Remark}
 \label{R1V16.2}
{\rm
We note that the gluing construction sketched below requires the smoothing operators of~\cite{ChDelayHilbert} when $4\le \ell< 6$.
}
\qed
\end{Remark}

\begin{proof} The proof is a repetition of that of~\cite[Proposition~3.2]{CCI2}, with the following minor changes arising because of different hypotheses and aims. First, there we use the Kerr
initial data outside and the data $(E,g,K)$ inside, while here  $(E,g,K)$ is outside while the family of initial data sets of Proposition~\ref{P1V16.1} is used inside. This change plays no role in the argument. Next, because of the different fall-off hypotheses, in the notation of~\cite{CCI2},  Equation~(10) there
is replaced now by
\begin{eqnarray}
 \label{afjest}
 \lefteqn{
    R \int\limits_{\{|\vec x|=1\}}\sum\limits_{j,k}
    (K^R_{jk}- (K^R)^{\ell}{}_\ell g^R_{jk})Y^j_i\nu^k d\sigma_e
    }
\\
    && = 
     R^{-1}\int\limits_{\{r=R\}}\sum\limits_{j,k}
    (K_{jk}-K^{\ell}{}_{\ell}g_{jk})Y^j_i\nu^k d\sigma_e
 =O (R^{1-2\alpha})
 \,.
  \nn
\end{eqnarray}
Here, replacing $\alpha$ by a smaller number if necessary, we have assumed that $\alpha<1$.
Similarly,~\cite[Equation~(11)]{CCI2} is replaced by
\begin{equation}
R \int\limits_{\{|\vec x|=1\}}
\Big[ \sum\limits_{i,j} x^\ell \left( g_{ij,i}-g_{ii,j}\right) \nu^j -
 \sum\limits_{i}\big(g_{ik}\delta^{k\ell} \nu^i- g_{ii}\nu^\ell \big) \Big] d\sigma_e
 =O (R^{1-2\alpha})
 \,.
 \label{afjest0}
\end{equation}
These estimates are obtained by straightforward adaptations of the argument following~\cite[Equation~(11)]{CCI2}.
The remaining arguments in~\cite{CCI2} remain unchanged.
\qed
 \end{proof}

\section{Existence of future-complete null hypersurfaces}
 \label{s2V16.3}

In this section we prove the \emph{``future-complete-null-hypersurfaces theorem''} for a large class of AE initial data.
We assume that the initial data have  well-defined center-of-mass and angular-momentum integrals. The general case will be settled in~\cite{ChBoost}.

\begin{Theorem}
  \label{T9V16.1}
  Let $(\hyp,g,K)$ be an $H^{-\alpha}_\ell$-AE  initial data set with $\alpha \ge 1/2 $, $\ell\ge 4$, and timelike four-momentum.
  Assume that  $(\hyp,g,K)$   is vacuum for sufficiently large distances, and that  \eq{15VIII16.1}-\eq{29IV16.2+} hold so that the center of mass and the angular momentum of $(\hyp,g,K)$ are well defined.
   Then the maximal globally hyperbolic development of $(\hyp,g,K)$ contains a \rfcp.
\end{Theorem}

\proof
For all $\varepsilon$ sufficiently small, consider the initial data
$$
 (\R^3\setminus B(2), g_\varepsilon, K_\varepsilon)
$$
obtained by scaling-down the
complement of a coordinate ball of radius $2/\varepsilon$ in an asymptotically Euclidean end of $(\hyp,g,K)$  defined, in local coordinates on $\R^3\setminus B(2)$,
as
\bea
 &
 g^\varepsilon_{ij} (\vec x):= g_{ij}(\vec x/\varepsilon)= \delta_{ij}
  + \varepsilon^{-\alpha}  o(|\vec x|^{-\alpha})
 \,,
  &
\\
 &
 K^\varepsilon_{ij} (\vec x):= K_{ij}(\vec x/\varepsilon)=
    \varepsilon^{-\alpha-1}  o(|\vec x|^{-\alpha-1})
 \,.
  &
\eeal{12V16.10}

Since $(\hyp,g,K)$ has   well defined center of mass and total angular momentum by hypothesis, Proposition~\ref{Pkerr} shows that for all $\varepsilon$ small enough the data
$(\R^3\setminus B(2), g_\varepsilon, K_\varepsilon)$ can be extended, by gluing, to a vacuum data set $(\R^3,\hat g_{\varepsilon},\hat K_{\varepsilon})$ with small weighted Sobolev norms as in point~\ref{p1V16.2} of Proposition~\ref{P1V16.1}.
Making  $\varepsilon$ smaller if necessary, the vacuum solution, say $(\calM,\fourg_\varepsilon)$ associated with the glued
initial data set will exist globally by Theorem~\ref{maintheoremlb2}. Uniqueness of solutions within domains of dependence guarantees that the space-time metric in the domain of dependence of $(\R^3\setminus B(4/\varepsilon),g,K)$ within the
space-time $(\mcM,\fourg)$ obtained by evolving $(\hyp,g,K)$ will, after a
constant rescaling of the space-time metric, be isometric to the domain of
dependence of $(\R^3,\hat g_{\varepsilon},\hat K_{\varepsilon})$ within  $(\calM,\fourg_\varepsilon)$. The result readily follows.
\hfill\qed

\section{The density argument}
 \label{s2V16.2}

Consider a  globally hyperbolic space-time  $(\mcM, \fourg)$  containing a family of outgoing null hypersurfaces $\mcN_u$ defined for all $u\in(-\infty, u_0]$ for some $u_0\in \R$. We suppose that those hypersurfaces have a well defined Trautman-Bondi mass $m (u)<\infty$
and that for all $u_1\le u_2\le u_0$ the mass-loss formula
\bel{5V16.11+}
 m (u_2) = m (u_1) - \int_{u_1}^{u_2}\int_{S^2} |N |^2 d\mu_0 du
\end{equation}
holds, where $|N |^2$ is the norm of the shear tensor, and $d\mu_0$ is the canonical measure on $S^2$.

We note that a version of \eq{5V16.11+} has been proved for the small-data space-times of Christodoulou and Klainerman, where $m (u)$ is the limit of the Hawking mass of a specific family of spheres constructed in~\cite{Ch-Kl}. Klainerman and Nicol\`o~\cite[Section~8.5]{KlainermanNicoloBook} proved a similar result under the same restrictive hypotheses on the asymptotics of the initial data as in~\cite{Ch-Kl} but without smallness restrictions.
By an abuse of terminology we will continue to call $m (u)$ the Trautman-Bondi mass in those contexts though, as already pointed out, the possibility of performing the steps needed for Bondi's construction of his mass in the category of space-times considered by~\cite{Ch-Kl} is far from clear.

One would like to prove that the limit
\bel{6V16.1}
 \mtot  :=\lim_{u\to-\infty} m (u)
\end{equation}
exists in $\R$, and  equals the ADM mass under rather general conditions.
Such a result has been proved in~\cite[Theorem~8.5.2]{KlainermanNicoloBook}
for  the class of space-times with initial data as in Definition~\ref{intSAFCK}. The existence of the limit \eq{6V16.1} has previously been established in~\cite{Ch-Kl}, where $\mtot $ has been called the \emph{total mass}, without relating $\mtot $ to the ADM mass.

Here we wish to point-out a density argument which would establish  the equality of the total mass with the ADM mass under rather general conditions. As a first step towards this, we note the following elementary consequence of \eq{5V16.11+}:

\begin{proposition}
 \label{P5VI16.1}
The limit \eq{6V16.1} exists in $\R\cup\{\infty\}$.
 We have
\bel{5V16.12}
 \mtot  < \infty
  \quad
   \Longleftrightarrow
   \quad
   \int_{-\infty}^{u_0}\int_{S^2} |N |^2 d\mu_0 du <\infty
   \,.
\end{equation}
If either of the two inequalities in \eq{5V16.12} is satisfied it holds that for all $u\le u_0$
\bel{5V16.13}
 m (u ) = \mtot   - \int_{-\infty}^{u}\int_{S^2} |N |^2 d\mu_0 du
 \,.
\end{equation}
\end{proposition}

\proof
The result is obtained by passing to the limit $u_1\to-\infty$ in \eq{5V16.11+}, using the monotone convergence theorem.
\qed

\medskip

To continue, let $(\mcM,\fourg)$ evolve  from an initial data $(\hyp,g,K)$ containing an asymptotically Euclidean end. For definiteness we assume that the data satisfy the vacuum constraint equations at sufficiently large distances. By~\cite{CorvinoSchoen2,ChDelay} there exists a sequence of initial data sets  $(\hyp,g_n,K_n)$ which satisfy the Klainerman-Nicol\`o conditions so that  $( g_n,K_n)$ converges to $(g,K)$, with $m_n$ tending to the ADM mass $m$ of $(\hyp,g,K)$ as $n$ tends to infinity.
 We have:

\begin{Proposition}
\emph{Assume} that the convergence of  $( g_n,K_n)$ to $( g ,K )$ is such that  we also have
 \label{P6V16.2}

\begin{enumerate}
 \item $N_n$ tends  to $N$ in $L^2((-\infty,u_0]\times S^2)$, and
 \item $m_n(u)$ tends pointwise to $m(u)$.
\end{enumerate}
Then the limit $\mtot =\lim_{u\to-\infty} m(u)$ exists, is finite, and equals the ADM mass $m$ of $(\hyp,g,K)$.

\end{Proposition}

\proof By~\cite[Section~8.5]{KlainermanNicoloBook} it holds that
\bel{5V16.11}
 m_n(u_2) = m_n(u_1) - \int_{u_1}^{u_2}\int_{S^2} |N_n|^2 d\mu_0 du
 \,.
\end{equation}
Since $m_n(u_1)$ tends to the (finite) ADM mass $m_n$ as $u$ tends to minus infinity,  by Proposition~\ref{P5VI16.1} we have, for all $u\le u_0$,
\bel{5V16.13+}
 m_n(u ) = m_n - \int_{-\infty}^{u}\int_{S^2} |N_n|^2 d\mu_0 du
 \,.
\end{equation}
%
%
Passing to the limit $n\to\infty$ gives
\bel{5V16.15}
 m (u )= \lim_{n\to\infty} m_n(u ) = \underbrace{\lim_ {n\to\infty} m_n}_{=m} - \int_{-\infty}^{u}\int_{S^2} |N|^2 d\mu_0 du
 \,.
\end{equation}
Passing with $u$ to $-\infty$ gives $\mtot  \equiv \lim _{u\to-\infty} m(u)=m$, as desired.
\qed

\medskip

We expect  the hypotheses of Proposition~\ref{P6V16.2} to be satisfied for a large class of space-times. It would be of interest to prove  precise statements to this effect.

\section{The  Trautman-Bondi mass in space-times with small slowly decaying initial data}
 \label{Structures}

It is a non-trivial fact that  the mass in the radiation regime for space-times as in Theorem~\ref{maintheoremlb2} can be defined by taking the limit, when receding to infinity in null directions, of the Hawking mass of suitable spheres. The proof of this requires the resolution of a certain amount of technical issues, which will be done in another paper~\cite{Bieriprep}.  The main difficulties arise from the borderline exponents $1/2$ and $3/2$ in \eq{afgenk}.  We note that one can also show that both the $\frac{1}{r}$ and  $\frac{1}{r^2}$  components of the Riemann tensor have a finite limit at null infinity with the borderline exponents.

It turns out that things become much simpler if decay rates larger than the $1/2$ and $3/2$ thresholds are assumed. This is the situation that will be discussed in this section.

We thus consider initial data of slightly stronger decay than in \cite{BieriPhD,BieriZipser1}:
{we will assume that there exists a coordinate system $(x^1, x^2, x^3)$ outside of a ball and a constant $\lalpha>0$ such that
\bea
 \label{22XI16.1}
 g_{ij} & = & \delta_{ij} + O_{H^3} (r^{-\frac{1}{2} - \lalpha})
\\
 K_{ij} & = & O_{H^2} (r^{-\frac{3}{2} - \lalpha})
 \,.
 \label{22XI16.2}
\end{eqnarray}
}
Clearly Theorem~\ref{maintheoremlb2} still applies for such data provided they are sufficiently small. However, we can use the extra $\lalpha$-decay in $r$ to say more about the solutions. Theorem~\ref{maintheoremlb2} holds with the change that our new $Q$ is slightly modified from the definition in (\ref{globalsalbQ}) by setting:
\bea
\label{NEWglobalsa1}
 \lefteqn{Q(a, x_{(0)})  =%
a^{-1}
\int_{{\hyp^3_0}}\ \big(\
(a^2 + d_0^2)^{\lalpha}
|K |^2
}
&&
\\ \nonumber
& &
\phantom{a^{-1} \int_{{\hyp^3_0}}\ }
    +  (a^2 + d_0^2)^{1+ \lalpha}  |\nabla K |^2
 + (a^2 + d_0^2)^{2 + \lalpha}  \ |\nabla^2 K |^2
\\
 \nonumber
 &&
 \phantom{a^{-1} \int_{{\hyp^3_0}}\ }
  +
(a^2 + d_0^2)^{1+ \lalpha}   |\mathrm{Ric}|^2
 + (a^2 + d_0^2)^{2 + \lalpha}    |\nabla \mathrm{Ric}|^2   \big)   d \mu_{g}
  \,.
  \nonumber
\end{eqnarray}
Let $\inf_{a, x_{(0)}}Q(a, x_{(0)})$ denote the infimum over all choices of origin $x_{(0)}$ and all $a$.
Then we consider asymptotically flat initial data sets with complete metric $\bar{g}$ and such that there exists a small positive $\epsilon_1$ such that
\be \label{NEWglobalsa}
 \inf_{a, x_{(0)} } Q(a, x_{(0)} ) < \epsilon_1  \,.
\end{equation}

{It follows from our results below that the behavior of the null asymptotics of these spacetimes is largely independent from the smallness assumptions.}
 Thus, the corresponding null asymptotics remains valid for spacetimes with large data. This is also true for the solutions constructed by Christodoulou-Klainerman in \cite{Ch-Kl} and for the ones by the first present author in \cite{BieriPhD,BieriZipser1}.

Let us introduce the foliation and its corresponding geometry that we will work with in this section.
Denote by $(\mcM,\fourg)$ the vacuum spacetime evolving from a set of initial data $(\hyp,g,K)$ satisfying
{\eqref{22XI16.1}-\eqref{22XI16.2},  with a  small $\epsilon_1$ in (\ref{NEWglobalsa})}. {This $\epsilon_1$ has to be suitably small depending on other quantities in order to close the proof, which consists of a bootstrap argument. In this argument we estimate quantities at times $t$ by their values in the initial hypersurface $\mcS_0$, which are controlled by $\epsilon_1$. The bootstrap assumptions in the spacetime slab require the considered quantities to be smaller than a small positive $\epsilon_0$. By choosing $\epsilon_1$ sufficiently small, these quantities will indeed become strictly smaller than $\epsilon_0$.

We choose to work with a maximal time function $t$ which foliates our spacetime into the $t$-level-sets which are maximal spacelike hypersurfaces $\mcS_t$. This means that the trace of the second fundamental form of each $\mcS_t$ is zero.   The spacetime $(\mcM,\fourg)$ is further} foliated by a function $u$ yielding as level sets the null hypersurfaces $\mcN_u$. The intersections
$$
 S_{t,u} = \hyp_t \cap \mcN_u
$$
are two-dimensional compact Riemannian manifolds. Consider a null frame $(f_1, f_2, f_3, f_4)$ with $f_4$ and $f_3$ denoting future-directed null vectors where $f_4$ is tangent to $\mcN_u$ as well as
\be
g(f_4, f_3) = -2
 \,,
\end{equation}
and an orthonormal frame $f_1$, $f_2$ on $S_{t,u}$.
The following scaling is useful in some arguments:
\[
(f_3, f_4) \mapsto (a^{-1} f_3, a f_4) \,, \ \ a > 0
\,.
\]
We call $r = r(t,u)$ the area radius of $S_{t,u}$, namely
$$
 r(t,u) = \sqrt{\frac{area(S_{t,u})}{4 \pi}}
  \,.
$$
Define
$$
 \tau_-^2 := 1 + u^2
 \,.
$$
The time vector field $T$ is defined as
$T =\frac{1}{2} (f_3 + f_4)$, whereas $N =  \frac{1}{2} (f_4 - f_3)$ is the outward normal to $S_{t,u}$ in $\hyp_t$.

Under the above assumptions, with respect to the foliation just introduced, a suitably tweaked version of the bootstrap arguments of~\cite{BieriPhD,BieriZipser1} shows that  the curvature components and main Ricci coefficients have the following behavior:
\bea
R_{A3B3} \ & = & \ \underline{\alpha}_{AB} = O(r^{-1} \tau_-^{-\frac{3}{2} - \lalpha})
 \,,
\label{intnullcurvalphaunderline*1} \\
R_{A334} \ & = & \ 2 \ \underline{\beta}_A  = O(r^{-2}  \tau_-^{-\frac{1}{2} -\lalpha})
 \,,
\\
R_{3434} \ & = & \ 4 \ \rho = O(r^{-\frac{5}{2} })
 \,,
 \\
\ ^* R_{3434} \ & = & \ 4 \ \sigma = O(r^{-\frac{5}{2} } \tau_-^{ -\lalpha})
 \,,
  \\
R_{A434} \ & = & \ 2 \ \beta_A  = o(r^{-\frac{5}{2} - \lalpha})
 \,,
 \\
R_{A4B4} \ & = & \ \alpha_{AB} = o(r^{-\frac{5}{2} - \lalpha})
 \,,
 \label{intnullcurvalpha*1} \\
\rho - \overline{\rho} \ & = & \ O(r^{-\frac{5}{2} } \tau_-^{ -\lalpha})
 \,,
  \\
\sigma - \overline{\sigma} \ & = & \ O(r^{-\frac{5}{2} } \tau_-^{ -\lalpha})
 \,.
\end{eqnarray}
Let $\{ S_s \}$ denote the affine foliation of $\mcN$, with $s$ the affine parameter function and $L$ the generating geodesic vector field of $\mcN$. Let $\underline{L}$ be the inward null normal. For $p \in S$ and $X, Y \in T_p S$ we define the second fundamental forms as
\bea
n (X, Y)  & = & g(D_X L, Y)  \label{n1}
 \,,  \\
\underline{n} (X, Y) & = & g(D_X \underline{L}, Y)
\,.
\label{n2}
\end{eqnarray}
(Note that $n$ and $\underline n$  are denoted by $\chi$ and by $\underline \chi$ in \cite{Ch-Kl}.)
We define the shears to be the traceless parts $\hat{n}$ respectively $\underline{\hat{n}}$ of the null second fundamental forms in (\ref{n1})- (\ref{n2}).

The shears behave as
\[
\hat{n} = O(r^{-\frac{3}{2}} \tau_-^{ - \lalpha} ) \,, \ \ \
\underline{\hat{n}} = O(r^{-1} \tau_-^{ - \frac{1}{2} - \lalpha} )
\]
The torsion
$$
t_A:= \frac 12  g(D_A L, \underline{L})
$$
(denoted by $\zeta_A$ in  \cite{Ch-Kl}) is of the order $O(r^{- \frac{3}{2}} \tau_-^{- \lalpha})$.

The second fundamental form $K$  {of the level sets of $t$}
 has the following components:
$K_{AB} = \eta_{AB}$, $K_{AN} = \epsilon_A$, $K_{NN} = \delta$.

We need  some more notation.
Let $V$ be a vector field tangent to $S$. Then we define the following norms on $S$:
\bea
\label{mainthmnormsS*2}
\parallel V \parallel_{p, S} (t,u) \ & = & \
\Big(
\int_{S_{t,u}} \mid V \mid^p \ d \mu_{\gamma} \Big)^{\frac{1}{p}}  \,, \ \ \mbox{ for } 1 \ \leq \ p \ < \infty
 \,,
\label{mainthmnormsS*1} \\
 \ & = & \
\sup_{S_{t,u}} \mid V \mid \ \  , \ \ \mbox{ for } p \ = \ \infty
 \,.
  \nn
\end{eqnarray}
Let $r_0(t)$ be the value of $r$ corresponding to the area of $S_{t,0}$, the surface of intersection between
$\mcN_0$ and $\hyp_t$, moreover $u_1(t)$ is the value of $u$ corresponding to $r_0(t)/2$.
We introduce the
interior and the exterior regions of each hypersurface $\hyp_t$.
The
\begin{itshape}interior region $I$\end{itshape},  denoted by $\hyp_t^i$,
consists of those points in $\hyp_t$ for which
\[
r \ \leq \ \frac{r_0(t)}{2} \,.
\]
The
\begin{itshape}exterior region $U$\end{itshape}, denoted by $\hyp_t^e$,
is defined as the collection of  points in $\hyp_t$ for which
\[
r \ \geq \ \frac{r_0(t)}{2} \,.
\]
Now, we introduce
\bea
\parallel V  \parallel_{p, i}  \ & = & \
\Big( \int_{\hyp_t^i} \mid V \mid^p \Big)^{\frac{1}{p}}  \,, \ \ \mbox{ for } 1 \ \leq \ p \ < \infty
\label{mainthmnormsI*1}  \,,
 \\
\parallel V  \parallel_{\infty, i}  \ & = & \
\sup_{\hyp_t^i} \mid V \mid  \,,
 \label{mainthmnormsI*2}  \\
\nonumber \\
\parallel V \parallel_{p, e} (t) \ & = & \
\Big( \int_{\hyp_t^e} \mid V \mid^p \Big)^{\frac{1}{p}}  \,, \ \ \mbox{ for } 1 \ \leq \ p \ < \infty
 \,,
\label{mainthmnormsU*1}  \\
\parallel V  \parallel_{\infty, e} (t)  \ & = & \
\sup_{\hyp_t^e} \mid V \mid  \label{mainthmnormsU*2}  \,.
\end{eqnarray}
We will see that each tensor behaves uniformly in the interior region, thus no specific properties of the components, whereas in the exterior region the components have different behavior. If we denote by $\nlap$ the intrinsic covariant derivative in $S_{t,u}$, then the following norms are controlled by small quantities:
\bea
r_0^{1 + \lalpha + q} \ \parallel D^q \ W \parallel_{2,i} \,, & & \\
\parallel \tau_-^{1 + \lalpha} r^q \nlap^q  \underline{\alpha} \parallel_{2, e}\,, & & \\
\parallel r^{1 + \lalpha + q} \nlap^q \alpha \parallel_{2, e}
\,,
& &  \\
\parallel \tau_-^{\lalpha} r^{q + 1} \nlap^q  \underline{\beta} \parallel_{2, e} \,, & &   \\
\parallel r^{1 + \lalpha + q} \nlap^q \beta \parallel_{2, e} & &  \\
\parallel \tau_-^{\lalpha} r^{q + 1} \nlap^q  (\rho - \overline{\rho}) \parallel_{2, e}  \,,& & \\
\parallel \tau_-^{\lalpha} r^{q + 1} \nlap^q  (\sigma - \overline{\sigma}) \parallel_{2, e}
 \,,& &
 \\
 r_0^{1 + \lalpha + q - \frac{2}{p}} \parallel D^q k \parallel_{p, i}
  \,,& & \\
 \parallel \tau_-^{\lalpha}  r^{(\frac{3}{2}  - \frac{3}{p} + q )} \ \nlap^q \delta \parallel_{p, e}
  \,, & &   \\
 \parallel \tau_-^{\lalpha} r^{(\frac{3}{2}  - \frac{3}{p} + q )} \ \nlap^q \epsilon \parallel_{p, e}
  \,, & &   \\
 \parallel \tau_-^{\lalpha} r^{(\frac{3}{2}   - \frac{3}{p} + q )} \ \nlap^q \hat{\underline{n}} \parallel_{p, e}
  \,, & &
\\
\parallel \tau_-^{\lalpha} r^{q} \ \nlap^q (\trace \,   n - \overline{\trace \,   n}) \parallel_{2, e}
 \,,
  & &   \\
\parallel  \tau_-^{\lalpha} r^{q} \ \nlap^q \hat{n} \parallel_{2, e}
 \,,
   & &   \\
\parallel  \tau_-^{\lalpha} r^{q} \ \nlap^q t \parallel_{2, e}
\,.
\end{eqnarray}
The corresponding norms are of course controlled as well on the null hypersurfaces $\mcN_u$.
We have only listed the norms relevant in the following sections. For more details see~\cite{Bieriprep}.

\subsection{Null Asymptotics}
 \label{ss23XI16.1}

By a straightforward modification of the proof in \cite{BieriPhD,BieriZipser1}, where now we assume a slightly stronger decay in $r$ respectively $u$ to the negative power of small, positive $\lalpha$, we derive the following asymptotic structures at future null infinity. On any null hypersurface $\mcN_u$ let $t \to \infty$. Thereby we compute the normalized curvature components $r\underline{\alpha}$ and $r^2 \underline{\beta}$. Note that the corresponding limits for $\alpha$ and $\beta$ are zero. Thus, we have
\bea
\lim_{\mcN_u, t \to \infty} r \underline{\alpha} & = & A(u, \cdot)
 \,,
   \\
\lim_{\mcN_u, t \to \infty} r^2 \underline{\beta} & = & B(u, \cdot)
 \,,
  \label{23XI16.1}
  \label{23XI16.3}
\end{eqnarray}
with
\bea
| A(u, \cdot) | & \leq & c (1 + |u|)^{- \frac{3}{2} - \lalpha}
 \,,
   \\
| B(u, \cdot) | & \leq & c (1 + |u|)^{- \frac{1}{2} - \lalpha}
 \,,
\end{eqnarray}
where $| \cdot |$ denotes the pointwise norm of the corresponding tensor on $S^2$ with respect to the standard metric on $S^2$.

{
 For completeness we list the key equations on which the derivation of the Bondi mass relies, from \cite{BieriPhD,BieriZipser1}. We refer to these works for details.}
{The null Codazzi equations read
\bea
\mydivergence \, \hat{n} \  - \  \frac{1}{2} \  d \ \trace \,    n \  + \hat{n} \cdot t  \ - \ \frac{1}{2} \ \trace \,   n \cdot t & = & - \beta \label{cl1}
 \,,
  \\
\mydivergence \, \hat{\underline{n}} \ - \ \frac{1}{2} \ d \ \trace \,   \underline{n} \ - \ \hat{\underline{n}} \cdot t \  + \ \frac{1}{2} \ \trace \,   \underline{n} \cdot t & = & \underline{\beta} \label{cl2}
\,.
\end{eqnarray}

The torsion $t$ obeys the system of equations
\bea
\curl  \ t \ & = & \  \sigma \ - \ \frac{1}{2} \  \hat{n} \wedge \underline{\hat{n}}
 \,,
  \\
\mydivergence \, t \ & = & \ - \mu \ - \ \rho \ + \ \frac{1}{2} \ \hat{n} \cdot \underline{\hat{n}}
\,.
\end{eqnarray}

The \emph{mass aspect functions} $\mu$ and $\underline{\mu}$ are defined as
\bea
\mu \ & = & \ K \ + \ \frac{1}{4} \ \trace \,   n \ \trace \,   \underline{n} \ - \ \mydivergence \, t  \label{ma1}
 \,,
  \\
\underline{\mu} \ & = & \ K \ + \ \frac{1}{4} \ \trace \,   n \ \trace \,   \underline{n} \ + \ \mydivergence \, t
 \,.
  \label{ma2}
\end{eqnarray}
with $K$ the Gauss curvature of $S_{t,u}$ and the Gauss equation
\be
K \ + \ \frac{1}{4} \ \trace \,   n \ \trace \,   \underline{n} \ - \ \frac{1}{2} \  \hat{n} \cdot \underline{\hat{n}} \ = \ - \rho
 \,.
 \label{K}
\end{equation}

Let $l = a^{-1} (T + N)$ and $\underline{l} = a (T - N)$. Also note that
\[
\frac{dr}{ds} \ = \ \frac{r}{2} \overline{\trace \,   n} \,.
\]
}

Using the propagation equation for $\hat{\underline{n}}$ one finds:

\begin{prop}
 \label{plimitsnnb}
On the null hypersurface $\mcN$ the limit
\[
\lim_{\mcN_u, t \to \infty} r \underline{\hat{n}} = N(u, \cdot)
\]
exists, with
\[
 {| N (u, \cdot) |_{\gamma} \leq c (1 + |u|)^{- \frac{1}{2} - \lalpha }
  \,.
}
\]
\end{prop}

\subsection{The mass}

{Let us sketch the proof, that in} the more general setting of spacetimes resulting from initial data as in \eqref{22XI16.1}-\eqref{22XI16.2}, {with the smallness condition  \eqref{NEWglobalsa}}, the Trautman-Bondi mass $m(u)$ for each null hypersurface $\mcN_u$ is well-defined and finite.   In~\cite{BieriPhD,BieriZipser1} the asymptotic behaviour in null directions is derived in a slightly more general setting. We use these results to establish Theorem~\ref{HBTM} below. Indeed, the situation investigated in \cite{BieriPhD,BieriZipser1} features different asymptotics due to the slow decay of the data. In particular, the usual peeling of the Weyl tensor is only established in two components.
However, there is enough structure at null infinity for the Bondi mass to have the desired properties.

Starting with formula (\ref{eq:mTB}) for the Trautman-Bondi mass, where $\mu$ is the mass aspect function,  we derive the corresponding formula for the situation with  {small data \eqref{22XI16.1}-\eqref{22XI16.2}}.  {It is proved in \cite{BieriPhD,BieriZipser1}  that the limit}
\be \label{malim1}
\lim_{\mcN_u, t \to \infty} r^3 \mu = \mu_L
\end{equation}
exists, and of course depends on $u$ and the angular variables.  Then the Trautman-Bondi mass $m(u)$ is defined as
\be \label{TBmL}
m(u) = \frac{1}{8 \pi} \int_{S^2} \mu_L d A_{S^2}
\end{equation}
with $d A_{S^2}$ denoting the area element of $S^2$. In \cite{BieriPhD,BieriZipser1}
the following propagation equation for $\mu$ is derived:
\be \label{propmu1}
\frac{\partial \mu}{\partial s} + \frac{3}{2} \trace \,   n \mu =
- \frac{1}{4} \trace \,   \underline{n} | \hat{n} |^2 + \frac{1}{2} \trace \,   n | t |^2 + 2 \mydivergence \, n \cdot t + \hat{n} \cdot \nabla \hat{\otimes} t
 \,.
\end{equation}
Thus we have:
\be \label{propmu2}
\frac{\partial \mu}{\partial t} = a \phi (
- \frac{1}{4} \trace \,   \underline{n} | \hat{n} |^2 + \frac{1}{2} \trace \,   n | t |^2 + 2 \mydivergence \, n \cdot t + \hat{n} \cdot \nabla \hat{\otimes} t )
 \,.
\end{equation}
{Here the function $a$ is the lapse function of the $u$-foliation,
%
\[
a = | d u |^{-1}
 \,,
\]
%
where the norm is taken with respect to the metric induced on the spacelike hypersurfaces $\hyp_t$.}

With the supplementary control on weighted $L^p$ norms on $S_{t,u}$ of the right-hand side we gain control on $r^3 \dan \mu \ \dan_{L^p(S_{t,u})}$.
Here the dimensionless $L^p$-norms on the surface $S_s$, for $p \geq 2$ is defined as:
\[
\dan \mu (s) \ \dan_{L^p(S)} \ = \
\big( \frac{1}{A(S_s)} \ \int_{S_s} \ \mid \mu  \mid^p_{\gamma} \ d \mu_{\gamma}  \big)^{\frac{1}{p}} \ = \
(4 \pi r^2)^{- \frac{1}{p}} \ \parallel \mu  \parallel_{L^p(S_s, \gamma)} \,.
\]

Let us sketch the proof that the Trautman-Bondi mass $m(u)$ tends to a limit $\mtot $ as $u \to  - \infty$.
In our notation, we write the Hawking mass as
\be
m_H (t,u) = \frac{r}{2} \big( 1 + \frac{1}{16 \pi} \int_{S_{t,u}} \trace \,   n \cdot \trace \,   \underline{n} \big)
 \,.
\end{equation}

Using equations (\ref{ma2})-(\ref{K}) and the Gauss-Bonnet formula, we find
\bea
\int_{S_{t,u}}\underline{\mu } &=&\int_{S_{t,u}}\left( \frac{1}{2}\widehat{%
n }\cdot \underline{\widehat{n }}-\rho \left( W\right) ) \right)
\label{integrate mu} \\
&=&4\pi \left( 1+\frac{1}{16\pi }\int_{S_{t,u}}\trace \,   n \cdot \trace \,   \underline{n }%
\right) =\frac{8\pi }{r}m_H
 \,,  \notag
\end{eqnarray}
\bea
\frac{\partial }{\partial t}m_H \left( t,u\right) &=&-\frac{r}{16\pi }%
\int_{S_{t,u}}\left( a\phi \trace \,   n - \overline{\phi a\trace \,   n }\right)
\underline{\mu }  \label{dt-hawking mass} \\
&&+\frac{r}{8\pi }\int_{S_{t,u}}a\phi \left( \frac{1}{2}\trace \,   n \left \vert
t \right \vert ^{2}-\frac{1}{4}\trace \,  \underline{n}\left \vert \widehat{%
n }\right \vert ^{2}\right) .  \notag
\end{eqnarray}

We observe that $K+\frac{1}{4}\trace \,   n \cdot \trace \,  \underline{n }$, and therefore $\underline{\mu }$, has the same decay properties in $r$ as in \cite{BieriPhD,BieriZipser1}. However, in the present situation  there is more decay in $\tau_-$. The results of \cite{BieriPhD,BieriZipser1} yield that
$m_H \left( t,u\right)$ has a finite limit as $t \to \infty$ on each fixed $\mcN_u$:
\be \label{limNEW}
\lim_{\mcN_u, t \to \infty} m_H(t,u) = m(u) \,.
\end{equation}
The precise estimate for the supplementary decay in $\tau_-$ is obtained as follows:
\bea
     \label{4Htmu}
& &
-
\int_0^{\infty}
\frac{r}{16\pi }%
\int_{S_{t,u}}\left( a\phi \trace \,   n -\overline{\phi a\trace \,   n }\right)
\underline{\mu }
\\
    &&+
\int_0^{\infty}
\frac{r}{8\pi }\int_{S_{t,u}}a\phi \left( \frac{1}{2}\trace \,   n \left \vert
t \right \vert ^{2}-\frac{1}{4}\trace \,  \underline{n }\left \vert \widehat{%
n }\right \vert ^{2}\right)   \notag \\
& \leq &
c \ \Big\{
-
\int_{\mcN_u} a
\frac{r}{16\pi }%
\left( a\phi \trace \,   n -\overline{\phi a\trace \,   n }\right)
\underline{\mu }  \nonumber \\
&&+
\int_{\mcN_u}
\frac{r}{8\pi } a^2 \phi \left( \frac{1}{2}\trace \,   n \left \vert
t \right \vert ^{2}-\frac{1}{4}\trace \,  \underline{n }\left \vert \widehat{%
n }\right \vert ^{2}\right)
\Big\}  \nonumber \\
& \leq &
c \tau_-^{- \lalpha} \parallel r \trace \,   n \parallel_{\infty, \mcN_u}
\parallel  r^{\frac{1}{2}} t \parallel_{L^2([0, t_*], L^4(S_{t,u})) } \nonumber \\
& & \ + \
c \tau_-^{- \lalpha}  \parallel r \trace \,   \underline{n} \parallel_{\infty, \mcN_u}
\parallel r^{\frac{1}{2}}  \hat{n} \parallel_{L^2([0, t_*], L^4(S_{t,u})) }    \nonumber \\
& & \ + \
c \tau_-^{- \lalpha}  \parallel r^{\frac{1}{2}} \mid   \trace \,   n - \overline{\trace \,   n}   \mid  \parallel_{L^2([0, t_*], L^4(S_{t,u})) }
\times
\parallel r^{3}  \underline{\mu}  \parallel_{L^2([0, t_*], L^4(S_{t,u})) }
 \,.
  \nn
\end{eqnarray}

Note that the boundedness of the terms on the right hand side does not depend on their behavior in $u$, but only on their decay properties in $r$. For the terms under consideration, the latter coincide with the situation in \cite{BieriPhD,BieriZipser1} and thus the terms on the right hand side of (\ref{4Htmu}) can be bounded using the results there.

Next, we   have
\begin{equation}
\frac{\partial }{\partial u}m_H \left( t,u\right) =\frac{1}{2}\overline{%
a\trace \,  \theta }m+\frac{r}{32\pi }\int_{S_{t,u}}a\left( \nabla _{N}\underline{\mu
}+\trace \,  \theta \underline{\mu }\right) .
\end{equation}
{The integral on the right-hand side of this equation is finite for the type of data that we are investigating here due to the extra $\lalpha$-decay in $u$ of $ {\underline{\mu}}$. Using the results in \cite{BieriPhD},  this integral can still be proved to be bounded  when $\lalpha=0$, but this requires much more work.}

Recall that $l=a^{-1}\left( T+N\right) $ and $\underline{l}=a\left( T-N\right) $.
Then it follows that
\begin{eqnarray*}
\nabla _{N}\underline{\mu }+\trace \,  \theta \underline{\mu } &=&\frac{1}{2}%
a^{2}\left( \mathbf{D}_{4}\underline{\mu }+\trace \,   n \underline{\mu }\right) -\frac{1}{2}\left( \mathbf{D}_{3}\underline{\mu }+\trace \,  \underline{n }%
\underline{\mu }\right)
 \,,
\end{eqnarray*}
and
\begin{eqnarray*}
\mathbf{D}_{4}\underline{\mu }+\trace \,   n \underline{\mu } &=&O(r^{- \frac{7}{2}}) \,,\\
\mathbf{D}_{3}\underline{\mu }+\trace \,  \underline{n }\underline{\mu } &=&-\frac{%
1}{4}\trace \,   n \left \vert \widehat{\underline{n }}\right \vert ^{2} +O(r^{- \frac{7}{2}})
\,.
\end{eqnarray*}
Hence,
\begin{equation*}
\frac{\partial }{\partial u}m_H \left( t,u\right) =\frac{r}{64\pi }%
\int_{S_{t,u}}\trace \,   n \left \vert \widehat{\underline{n }}\right
\vert ^{2}  +O\left( r^{- \frac{1}{2}}\right) .
\end{equation*}%

{The following has been proved in~\cite{BieriZipser1}:} (1) the metric $\widetilde{\gamma }=\phi _{t,u}^{\ast }\left(
r^{-2}\gamma \right) $ converges to the standard metric $\overset{\circ }{%
\gamma }$ of the unit sphere $S^{2}$ as $t\rightarrow \infty $ for each $u$ (%
$\phi _{t,u}^{\ast }$ is a diffeomorphism from $S^{2}$ to $S_{t,u}$), (2) $%
\frac{r}{2}\trace \,   n $ converges to $1$, and (3) $r\widehat{\underline{n }}$
converges to $-2\Xi $.
Moreover, the limit of the derivative of the local quantity on the left hand side in the previous formula is indeed the derivative of the Bondi mass in (\ref{Bmassloss}).
Using this one obtains the following \emph{Bondi-mass-loss-type} formula:
\begin{equation} \label{Bmassloss}
\frac{\partial }{\partial u} m \left( u\right) =\frac{1}{8\pi }%
\int_{S^{2}} \left \vert N (u, \cdot ) \right \vert ^{2} d\mu _{\overset{\circ }{\gamma }}
 \,.
\end{equation}%
(In our context the property, that the derivative of the limit is  the limit of the derivative, is non-trivial  but was essentially resolved in \cite{Ch-Kl}, beginning of chapter 17, in particular see  p.\ 493 (17.0.1), and p.\ 494.)

From the results of Proposition~\ref{plimitsnnb} on $N(u, \cdot)$ it is clear that the right hand side of (\ref{Bmassloss}) is integrable in $u$. As it is also positive, the Trautman-Bondi mass m(u) is a non-increasing function of $u$ taking values from its finite limit $\mtot $ as $u \to - \infty$ to its finite limit $m_0$ as $u \to \infty$. Thus, it is $0 \leq m_0 \leq \mtot $, where the first inequality is due to the positive mass theorem. If the spacetime was constructed under smallness assumptions on the initial data such as in the works of \cite{Ch-Kl} or \cite{BieriPhD,BieriZipser1}, one can show that $m_0 = 0$, thus all the mass would be radiated away. This can be seen from equation (\ref{integrate mu}): We take the limit for fixed $u$ as $t \to \infty$ and use the results on the null asymptotic behavior of the involved quantities. It then follows from the smallness assumptions on the data in the works just cited that $m_0  =  0$. However, in general without such smallness assumptions this $m_0$ is the mass left in the system after radiation.

The above discussion can be summarised as:

\begin{The} \label{HBTM}
For spacetimes resulting from evolved initial data as in our \eqref{22XI16.1}-\eqref{22XI16.2} and under our assumptions (\ref{NEWglobalsa}) the following holds:
\begin{enumerate}
\item[(a)] The Hawking mass $m_H (t,u)$ tends to the Trautman-Bondi mass $m(u)$ on each $\mcN_u$:
\[
\lim_{\mcN_u, t \to \infty} m_H (t,u) = m (u)
 \,.
\]
\item[(b)] The Bondi mass loss formula holds
\[
\frac{\partial }{\partial u} m \left( u\right) =\frac{1}{8\pi }%
\int_{S^{2}} \left \vert N (u, \cdot ) \right \vert ^{2} d\mu _{\overset{\circ }{\gamma }}
 \,.
\]
\item[(c)] The Trautman-Bondi mass $m(u)$ tends to a limit $\mtot$ as $u \to - \infty$.
\end{enumerate}
\end{The}

The last statement follows from the above and the monotone convergence theorem.

We finally note~\cite{Bieriprep}:

\begin{theorem} \label{TBADM}
The total mass $\mtot$ of Theorem~\ref{HBTM} equals the ADM mass.
\end{theorem}

\appendix

\section{Notations}
 \label{s19V16.1}

For $\alpha\in \R$ and $k\in \N$  we define
\bean
 \lefteqn{
  C^\alpha_k :=\{u\in C^k(\R^3) \ \mbox{such that}
   }
  &&
\\
&&
 \mbox{for}\   0\le \ell \le k\
  \mbox{we have}\
\sup_{\R^3} |\nabla ^\ell u\normwithnodelta   (1+r)^{-\alpha} < \infty
\}
 \,,
\eeal{5V16.9}
with the obvious norm. Here $|\nabla^\ell u\normwithnodelta  $ is the Euclidean norm of the tensor of covariant derivatives with respect to the flat metric $\delta$ of a tensor $u$, with $r=|\vec x\normwithnodelta  $.

Given an open set $\Omega\subset \R^n$, for $p\ge 1$, $\alpha\in \R$ and $k\in \N$ we set
\bea
 \label{12V16.1}
 \lefteqn{
  W^\alpha_{k,p}(\Omega) :=\Big\{\mbox{the completion of $C^\infty_0(\Omega) $  with respect to}
   }
  &&
\\
\nn
&&
 \|u\|^p_{ W^\alpha_{k,p}} := \sum_{0\le \ell \le k}\int_{\Omega} (1 +|\vec x|)^{-p\alpha - n }|\nabla ^\ell u|^p_\delta \, d^nx
\Big\}
 \,.
\eea
We write  $W^\alpha_{k,p}$ when $\Omega=\R^n$ or is otherwise obvious from the context. For $\ell<k-n/p$ we have the compact inclusion $W^{\alpha}_{k,p}\subset C^\alpha_\ell$ and in fact for large $r\equiv |\vec x|$ it holds that (cf., e.g.,~\citep{Bartnik86})
\bel{12V16.2}
u \in W^{\alpha}_{k,p}\,,
 \
 \ell<k-n/p
 \quad
 \Longrightarrow
 \quad
  |\nabla^\ell u\normwithnodelta   = o(r^{\alpha})
  \,.
\end{equation}
We set
\bel{19V16.2}
 H^\alpha_k := W^{\alpha}_{k,2}
 \,.
\end{equation}

We shall write
\bel{19V16.1}
 u=O_{H^k}(r^\alpha)
\end{equation}
if the norm $\|u\|_{H^\alpha_k}$ of the tensor field $u$ is finite.
With this notation, the implication \eq{12V16.2} with $p=2$ can  be rewritten as
\bel{12V16.2+}
  u=O_{H^k}(r^\alpha)\,, \ k> \ell+n/2
 \quad
 \Longrightarrow
 \quad
  |\nabla^\ell u\normwithnodelta   = o(r^{\alpha})
  \,.
\end{equation}

\section{Boosting Schwarzschild}
 \label{A3VI16.1}

Consider the Schwarzschild metric with mass $m$, which we denote by $^4g_m$, in isotropic coordinates,
\bel{3VI16.1}
 ^4 g_m= -A(s) dt^2+ B(s) (dx^2 + dy^2 +dz^2)
 \,,
\end{equation}
where
\bel{3VI16.2}
 \quad s:= \frac m r
 \,,
 \quad
 A(s) =\left(\frac{ 1 - \frac s 2}{1+\frac s 2}\right)^2 \approx 1 - 2 s
 \,,
 \quad
 B(s) = \left(1 + \frac s 2\right)^4 \approx 1 + 2 s
 \,.
\end{equation}
Let us introduce new coordinates $(\ol t, \ol x, \ol y, \ol z)$ defined as
\bel{3VI16.3}
 t = \gamma(\ol t - v \ol z)
 \,,
  \
 x= \ol x
 \,,
  \
 y= \ol y
 \,,
 \
 z = \gamma(\ol z - v \ol t)
 \,,
\end{equation}
where $\gamma$ is the usual boost factor in units where $c=1$, $\gamma^{-2}:=1-v^2$.
In these coordinates the metric becomes
\bean
 ^4 g_m &=& -\gamma^2(A - v^2 B) d\ol t^2 + 2 v \gamma^2  (A-B) d\ol t d \ol z
 + B (d\ol x^2 + d\ol y^2) + \gamma^2 (B - v^2 A) d\ol z^2
\\
 & = & -\big(1 - 2 \gamma^2(1+v^2) s + O(s^2)\big)d\ol t^2 - 8  v \gamma^2s \big(1+O(s)\big) d\ol t d\ol z
 \nn
\\
 &&
  + \big(1+2 s +O(s^2)\big) (d\ol x^2 +d\ol y^2) + \big(1+2 \gamma^2 (1 + v^2) s+O(s^2)\big) d\ol z^2
 \,.
\eeal{3VI16.4}
Here all the error terms $O(s^i)$ are smooth functions of $s$ near $s=0$,
and of course
\bel{3VI16.5}
 s =\frac m r = \frac m { \sqrt{\ol x^2 + \ol y^2 + \gamma^2 (\ol z - v \ol t)^2} }
 \,.
\end{equation}
Let $\ol n=\ol n^\mu \partial_\mu = \ol n^{\ol t} \partial_{\ol t} + \ol n^{\ol z} \partial_{\ol z}$ denote the future-directed unit normal to the level sets of $
\ol t$, thus
\bel{3VI16.6}
 0= \ol n_{\ol z} = g_{\ol z \ol t} \ol n ^{\ol t} +
 g_{\ol z \ol z} \ol n ^{\ol z}
 \quad
  \Longrightarrow
  \quad
  \ol n = \big(1+O(s)\big) \partial_{\ol t}  - 4  v \gamma^2s \big(1+O(s)\big)  \partial_{\ol z}
   \,.
\end{equation}
The extrinsic curvature tensor, which we denote by $ K_p$, of the level set $\ol t=0$ therefore reads
\bea
\label{3VI16.7}
  K_p & = & \mcL_{\ol n} g _{ij}  d\ol x^i d\ol x^j
  =
 \big(\ol n^\mu \partial_\mu g_{ij}
 + \partial_i \ol n^\mu g_{j \mu}
 + \partial_j \ol n^\mu g_{i \mu}
 \big) d\ol x^i d\ol x^j
\\
 & = &
 \big(\ol n^{\ol t} \partial_s g_{ij} \frac{\partial s}{\partial \ol t}
 + \partial_i \ol n^{\ol z} \delta_{j z}
 + \partial_j \ol n^{\ol z} \delta_{i z}
  + r^{-2}O(s)
 \big) d\ol x^i d\ol x^j
 \nn
\\
 & = &
 \big(  \partial_s g_{ij} \frac{\partial s}{\partial \ol t}
  + r^{-2}O(s)
 \big) d\ol x^i d\ol x^j
 + 2 \partial_s \ol n^{\ol z} \frac{\partial s}{\partial \ol x^i } d\ol x^i d\ol z
 \nn
\\
 & = &
   \frac{2m\gamma^2 v \ol z}{r^3}
    \big(d\ol x^2 +d\ol y^2+ \gamma^2(1+v^2)  d\ol z^2\big)
 \nn
\\
 &&
 -8v \gamma^2 \frac{m}{  r^3} (\ol x d \ol x + \ol y d \ol y + \gamma^2 \ol z d\ol z) d\ol z
  + r^{-2}O(s)
  d\ol x^i d\ol x^j
 \,.
  \nn
\eea
Let $g_p$ denote the metric induced on the level sets of $\ol t$, and let $(g_{p,\vec a},K_{p,\vec a}$ be the Riemannian metric and the extrinsic curvature tensor obtained by applying a translation by a vector $-\vec a$ to $(g_p,K_p)$. Setting
$$
 s_{\vec a} \equiv \frac m {r_{\vec a}} :=\frac m { \sqrt{(\ol x - a_1)^2 +(\ol y - a_2)^2 + \gamma^2 (\ol z - a_3)^2}}
 \,,
$$
we find
\bea
  \label{5VI16.2}
  g_{p,\vec a} &=&
   \big(1+2 s_{\vec a} +O(s^2_{\vec a})\big) (d\ol x^2 +d\ol y^2)
\\
 &&   + \big(1+2 \gamma^2 (1 + v^2) s_{\vec a}+O(s_{\vec a}^2)\big) d\ol z^2
  \nn
 \,,
\\
 \label{5VI16.1}
  K_{p,\vec a}
 & = &
   \frac{2m\gamma^2 v (\ol z-a_3)}{r^3_{\vec a}}
    \big(d\ol x^2 +d\ol y^2+ \gamma^2(1+v^2)  d\ol z^2\big)
\\
 \nn
 &&
 -8v \gamma^2 \frac{m}{  r^3_{\vec a}} \big((\ol x-a_1) d \ol x + (\ol y-a_2) d \ol y + \gamma^2 (\ol z - a_3) d\ol z\big) d\ol z
\\
 \nn
 &&
  + r^{-2}_{\vec a}O(s_{\vec a})
  d\ol x^i d\ol x^j
 \,.
\eea
We note that at $\ol t$ the metric is even and the extrinsic curvature tensor is odd.

\bigskip

\noindent{\sc Acknowledgements:} We are grateful to the Center for Mathematical Sciences and Applications at Harvard University for hospitality and support during part of work on this paper.  This work has been further supported in part by the Austrian Science Fund (FWF)  project P29517-N16.  {The first author is grateful to Demetrios Christodoulou who provided the main ideas for the analysis of the mass in her thesis.}

\bibliographystyle{amsplain}
\bibliography{%
../references/reffile,%
../references/newbiblio,%
../references/hip_bib,%
../references/newbiblio2,%
../references/bibl,%
../references/howard,%
../references/bartnik,%
../references/myGR,%
../references/newbib,%
../references/Energy,%
../references/netbiblio,%
../references/PDE}

\end{document}